\documentclass[12pt]{article}
\usepackage{graphicx}
\usepackage{epsfig}
\begin {document}
\begin{center}
\bf
STRING JUNCTION AND DIFFUSION OF
BARYON CHARGE IN MULTIPARTICLE
PRODUCTION PROCESSES

\bigskip

Yu. M. Shabelski

\vskip 0.5 truecm

Petersburg Nuclear Physics Institute, Gatchina, St. Petersburg, Russia\\

\vskip 1. truecm

{\it Lecture given at XXXX PNPI Winter School of Physics, Repino, St.Petersburg, February 2006}

A b s t r a c t
\end{center}

We consider the phenomenological consequences of the assumption that
the baryons are the systems of three quarks and string junction.
The process of baryon number transfer due to string junction propagation
in rapidity space is considered in detail. At high energies it leads to
a significant effect in the net baryon production in $hN$ and
$hA$ collisions at mid-rapidities and in the incident meson
fragmentation region. The results of numerical calculations in the
framework of the Quark--Gluon String Model are in reasonable agreement
with the data.

\vskip 1cm


\vskip 3.5 truecm
E-mail: shabelsk@thd.pnpi.spb.ru

\newpage

\section{Introduction}

The Quark--Gluon String Model (QGSM) and the Dual Parton Model (DPM) are
based on the Dual Topological Unitarization (DTU) and describe quite
reasonably many features of high energy production processes, in both
hadron--nucleon and hadron--nucleus collisions \cite{KTM}--\cite{Sh1}.
High energy interactions are considered as going via the exchange
of one or several Pomerons, and all elastic and inelastic processes
result from cutting through or between Pomerons \cite{AGK}. Inclusive
spectra of hadrons are related to the corresponding fragmentation
functions of quarks and diquarks, which are constructed using the
Reggeon counting rules \cite{Kai}.

In the present lecture, we discuss the processes connected with the
transfer of baryon charge over long rapidity distances. In the string
models, baryons are considered as configurations consisting of three
strings (related to three valence quarks) connected at the point called
``string junction" (SJ) \cite{Artru}--\cite{Khar}. The string junction
has a nonperturbative origin in QCD. Many phenomenological results were
obtained 25 years ago \cite{IOT,RV}, \cite{IOT1}--\cite{Noda}.
They are discussed shortly in Sect.~2.

It is very important to understand the role of the string junction
in the dynamics of high-energy hadronic interactions.
Now we have several different experimental results concerning
the processes of baryon charge transfer.

The most impressive are the data \cite{ait} on $\Omega/\bar{\Omega}$
asymmetry in collisions of particles without strange quarks in the
initial state.  This asymmetry is absent in many phenomenological
models. It appears as a result of SJ existing in the initial state and
it was predicted qualitatively in \cite{Noda}.

The data \cite{Bren} clearly show that in the forward hemisphere the
number of secondary protons produced in $\pi^+p$ interactions is
significantly larger than the number of secondary antiprotons produced
in $\pi^-p$ collisions. This difference cannot be described
\cite{Sh} without the assumption that the baryon charge is
transferred from the target proton to the pion hemisphere.

There exist the data obtained on nuclear targets. The discussed
effects are confirmed by the measurements of $\Lambda/\bar{\Lambda}$
and $\Xi/\bar{\Xi}$ hyperon production asymmetries in 500\,GeV/c
$\pi^-$-nucleus interactions \cite{ait}. Similar data on the
differences of $p-\bar{p}$ yields in $\frac12(\pi^+p+\pi^-p)$
collisions together with the same data for $Pb$ target at 158\,GeV/c
were presented by the NA49 Coll. \cite{NA49}. The data \cite{WA89} on
hyperon production by $\pi^-$ beam on carbon and copper targets
show the evident and numerically large difference in the spectra
of secondary baryons and antibaryons.

A second group of data concerns the energy dependence of the
differences in yields of the protons and antiprotons at 90$^\circ$
({\em i.e.} at $x_F=0$) at ISR energies \cite{ISR}. Similar data
on $\bar{p}/p$ ratio at RHIC energies \cite{BRA,RHIC} were published
recently. Finally, the proton--antiproton asymmetry in photoproduction
was measured at HERA \cite{H1}.

Quantitative theoretical description of baryon number transfer via
string junction mechanism was suggested in 90's. In \cite{KP1}, the
experimentally observed $p/\bar{p}$ asymmetry at HERA energies was
predicted that was confirmed by the data \cite{H1} which were obtained
later. This asymmetry was considered in more detail in \cite{KP2,GKP}.
In \cite{Bopp}, it was noted that the $p/\bar{p}$ asymmetry measured at
HERA can be obtained by simple extrapolation of ISR data.

The important theoretical results on the baryon number transfer due to
SJ diffusion in rapidity space were obtained in \cite{ACKS} and
following papers \cite{SJ1}--\cite{Olga}. In the present lecture, we
consider the main results for the case of secondary baryon/antibaryon
production both from nucleon and nuclear targets.

The most interesting are the differences in baryon and antibaryon
production in the meson beam fragmentation. In the case of nuclear
target the discussed effects are enhanced due to two reasons. First,
the usual production of secondaries (which can be considered as a
background for string junction effects) in the beam fragmentation
region is suppressed due to nuclear absorption \cite{KTMS},
  [37--39].
Second, the probability of the baryon number transfer should be
proportional to the number of inelastic interactions in the nuclear
matter, $\langle\nu\rangle_{hA}$.

In the case of baryon beams the SJ effects are the most important in
the central (midrapidity) region \cite{ACKS,SJ1}.

\section{Baryon as \boldmath$3q+SJ$ system}

In QCD, the hadrons are composite bound state configurations built up
from the quark $\psi_i(x), i = 1,...N_c$ and gluon $G^\mu_a(x)$,
$a=1,...,N_c^2-1$ fields. In the string models the meson wave
function has the form of ``open string" \cite{Artru,RV}, as is
shown in Fig.~1a.

\begin{figure}[htb]
\centering
\includegraphics[width=.5\hsize]{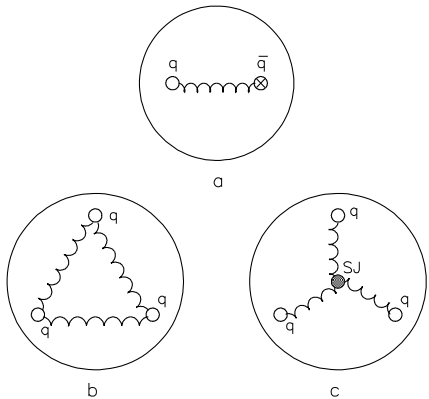}
\vskip -.3cm
\caption{\footnotesize
Composite structure of a meson (a) and baryon (b) and (c)
in string models. Quarks are shown by open points and antiquarks by
crossed points}
\end{figure}

The meson wave function reads as follows:
\begin{eqnarray}
&& M\ =\ \bar{\psi}^i(x_1) \Phi_i^{i'}(x_1,x_2)\psi_{i'}(x_2)\,,
\\
&& \Phi_i^{i'}(x_1,x_2) = \left[ T\exp \left(g \int\limits_{P(x_1,x_2)}
A_{\mu}(z) dz^{\mu}\right) \right]_i^{i'}.
\end{eqnarray}

In the last equation $P(x_1,x_2)$ represents a path from $x_1$ to
$x_2$ which looks like an open string with ends in $x_1$ and $x_2$.

For the baryons there exist two possibilities, ``triangle", or
$\Delta$ connection shown in Fig.~1b and ``star", or $Y$ connection
shown in Fig.~1c. The last variant is considered as the most
interesting. Here a baryon is considered as configurations consisting
of three strings attached to three valence quarks and connected in a
point called the ``string junction" (SJ) \cite{Artru,RV}. The
correspondent wave function can be written as
\begin{eqnarray}
&&B\ =\ \psi_i(x_1) \psi_j(x_2)\psi_k(x_3) J^{ijk}\,,
\\
&& J^{ijk} =\ \Phi^i_{i'}(x_1,x)\Phi_{j'}^j(x_2,x)\Phi^k_{k'}(x_3,x)
\epsilon^{i'j'k'} \,.
\end{eqnarray}

Such baryon wave function can be defined as a ``star" or ``Y" shape
and it is preferable \cite{Artru,RV} in comparison with ``triangle"
(``ring") or ``$\Delta$" shape, where the problems with gauge
invariance appear \cite{RV}. This ``Y" structure of baryon is confirmed
by lattice calculations \cite{latt}.

The presented picture leads to several phenomenological predictions.

In particular, there exist the rooms for exotic states, such as
glueball, or gluonium (``closed string"), Fig.~2a, \cite{RV,Ven}
\begin{equation}
\mbox{Glueball  = Tr} \left[T\exp \left(g \int\limits_{P(closed)}
A_\mu(z) dz^\mu\right) \right].
\end{equation}

\begin{figure}[htb]
\centering
\vskip -1cm
\includegraphics[width=.5\hsize]{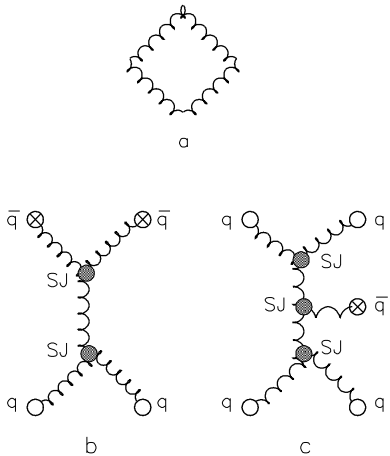}
\vskip -.5cm
\caption{\footnotesize
Exotic states: glueball (a), 4-quark meson (tetraquark)
$M_4 = qq\bar{q}\bar{q}$ (b) and 5-quark baryon (pentaquark)
$B_5 = qqqq\bar{q}$ (c) in string models. Quarks are shown by open
points and antiquarks by crossed points}
\end{figure}

The multiquark bound states, such as 4-quark meson, Fig.~2b, pentaquark,
Fig.~2c, {\em etc.} also can exist \cite{RV}, \cite{DPP1}--\cite{DPP3}.
Without specifying the model it is impossible to say anything definite 
about the sign of the correspondent binding energy, {\em i.e.} are they 
the bound states or not. However we can expect that the part of a 
particle momentum carried out by gluons in the case of multiquark states 
should be larger than for usual particles, Figs. 1a and 1c, due to the 
larger number of string junctions.

From the point of view of the Additive Quark Model, a meson consists,
as before, of two constituent quarks, Fig.~1a, but a baryon
consists now of four constituent objects, three constituent
quarks and SJ, as is shown in Fig.~1c. In such a picture the
ratio of nucleon--nucleon and meson--nucleon total cross sections
at high energies increases \cite{KKKY} as compared to 
classical result \cite{LF}
$\sigma(N-N)/\sigma(\pi-N)=3/2$, namely, with accounting for
the possibility of SJ interaction with a target
\begin{equation}
\frac{\sigma (N-N)}{\sigma (\pi-N)}\ =\ \frac32 +
\frac{\sigma (SJ-N)}{2\sigma(q-N)}\,,
\end{equation}
where the additional term $\sigma(SJ-N)/(2\sigma(q-N))$ can be
estimated \cite{KKKY} to equal $1/4\div1/6$. This correction
results in a better agreement \cite{AKNS} with experimental data.

The $B\bar{B}$ annihilation cross section, $\sigma_{ann}$, is not
necessarily equal to the difference
$\Delta \sigma=\sigma^{tot}(B\bar{B})-\sigma^{tot}(BB)$ \cite{RV}.

The existence of SJ in a baryon structure changes the quark counting 
rules for reactions with large momenta transfer \cite{IOT1,Noda}. The
reaction $\bar{p}p\to\bar\Omega\Omega$ can occur now without
breaking the OZI rules. The ratio of $\bar\Omega/\Omega$ production
for the collisions of non-strange hadrons is predicted to be smaller
than unity \cite{Noda} contrary to many models for multiparticle
production. This prediction is in agreement with the experimental data
\cite{ait} and their model description in \cite{ACKS,SJ1}.

In the case of inclusive reactions the baryon number transfer to large
rapidity distances in hadron--nucleon reactions can be explained by SJ
diffusion \cite{KP1,ACKS}. Now we consider several examples of effects
connected with the string junction diffusion for hadron--nucleon and
hadron--nucleus inelastic interactions.

\section{Inclusive spectra of secondary hadrons \newline in the
Quark--Gluon String Model}

For more quantitative predictions we need certain model for
multiparticle production and we will use the QGSM for the numerical
predictions presented below.

As was mentioned above, the high energy hadron--nucleon and 
hadron--nucleus interactions are considered in the QGSM as going via the 
exchange of one or several Pomerons. Each Pomeron corresponds to a 
cylindrical diagram, see Fig.~3a, thus, when cutting a Pomeron, two 
showers of secondaries are produced as is shown in Fig.~3b. The 
inclusive
spectrum of secondaries is determined by the convolution of diquark,
valence quark and sea quark distributions $u(x,n)$ in the incident
particles and the fragmentation functions $G(z)$ of quarks and diquarks
into secondary hadrons. The diquark and quark distribution functions
depend on the number $n$ of cut Pomerons in the considered diagram.

\begin{figure}[htb]
\centering
\includegraphics[width=.6\hsize]{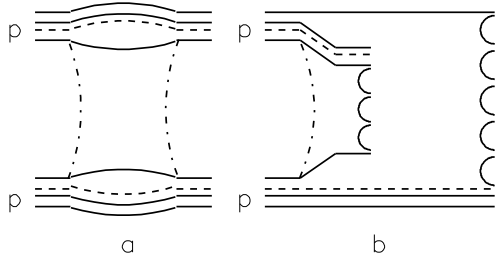}
\vskip -.5cm
\caption{\footnotesize
Cylindrical diagram corresponding to the
one--Pomeron exchange contribution to elastic $pp$
scattering (a) and its cut which determines the contribution to
inelastic $pp$ cross section (b). Quarks are shown by solid curves and
 SJ by dashed curves}
\end{figure}

For nucleon target, the inclusive spectrum of a
secondary hadron $h$ has the form \cite{KTM}:
\begin{equation}
\frac x{\sigma_{inel}} \frac{d\sigma}{dx}\ =\ \sum_{n=1}^\infty
w_n\phi_n^h (x)\ ,
\end{equation}
where the functions $\phi_{n}^{h}(x)$ determine the contribution of
diagrams with $n$ cut Pomerons and $w_n$ is the probability of
this process. Here we neglect the contributions of diffraction
dissociation processes which are comparatively small in the majority of
processes considered below. They can be separately accounted for
\cite{KTM,2r,Sh}.

For $pp$ collisions
\begin{eqnarray}
\phi_{pp}^h(x) &=& f_{qq}^{h}(x_+,n)f_q^h(x_-,n) +
f_q^h(x_+,n)f_{qq}^h(x_-,n)
\nonumber\\
&&\hspace*{3.5cm}+\ 2(n-1)f_s^h(x_+,n)f_s^h(x_-,n)\ ,
\\
x_{\pm} &=& \frac12\left[\sqrt{4m_T^2/s+x^2}\ \pm x\right] ,
\end{eqnarray}
where $f_{qq}$, $f_q$ and $f_s$ correspond to the
contributions of diquarks, valence and sea quarks respectively.

The last functions are determined by the convolution of the diquark and
quark distributions with the fragmentation functions, for example,
\begin{equation}
f_q^h(x_+,n)\ =\ \int\limits_{x_+}^1u_q(x_1,n)G_q^h(x_+/x_1) dx_1\ .
\end{equation}

For meson beam the diquark contribution $f^h_{qq}(x_+,n)$
in Eq. (8) should be changed by the contribution of valence antiquarks:
\begin{equation}
\phi_{\pi p}^h(x) = f_{\bar{q}}^h(x_+,n)f_q^h(x_-,n) +
f_{q}^{h}(x_{+},n)f_{qq}^{h}(x_{-},n) +
2(n-1)f_{s}^{h}(x_{+},n)f_{s}^{h}(x_{-},n)\ .
\end{equation}

The diquark and quark distributions as well as the fragmentation
functions are determined by Regge intercepts.

According to \cite{ACKS,SJ1}, we consider three possibilities to obtain
the net baryon charge. The first one is the fragmentation of the
diquark giving rise to a leading baryon (Fig.~4a). A second
possibility is to produce a (leading) meson in the first break-up
of the string and baryon in the subsequent break-up (Fig.~4b).
In the considered approach the baryon consists of three valence quarks
together with string junction (SJ), which is conserved during the
interaction.\footnote{At very high energies one or even several SJ pairs
can be produced.} This gives a third possibility for secondary baryon
production in non-diffractive hadron--nucleon interactions, shown in
Fig.~4c.

\begin{figure}[htb]
\centering
\includegraphics[width=.55\hsize]{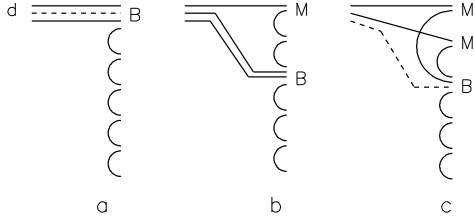}
\vskip -0.1cm
\caption{\footnotesize
QGSM diagrams describing secondary baryon $B$ production
by diquark $d$: initial SJ together with two valence quarks and
one sea quark (a), together with one valence quark and two sea
quarks (b) and together with three sea quarks (c)}
\end{figure}

In Fig.~4a, the secondary baryon consists of the SJ together with
two valence ($qq$) and one sea ($s$) quarks, in Fig.~4b of one valence
and two sea quarks and in Fig.~4c of three sea quarks. The corresponding
fragmentation functions for the secondary baryon $B$ production can be
written as follows (see \cite{ACKS,SJ1} for more detail):
\begin{eqnarray}
G^B_{qq}(z) &=& a_N v_{qq} z^{2.5} \;,
\\
G^B_{qs}(z) &=& a_N v_{qs} z^ 2 (1-z) \;,
\\
G^B_{ss}(z) &=& a_N \varepsilon v_{ss} z^{\alpha_{SJ}-1} (1-z)^2
\end{eqnarray}
for the processes shown in Figs.~4a, 4b and 4c, respectively. The
fraction $z$ of the incident baryon energy carried by the secondary
baryon increases from Fig.~4a to Fig.~4c, whereas the mean rapidity
gap between the incident and secondary baryon increases. In
Eqs.~(12)--(14), $a_N$ is the normalization parameter for secondary
baryon production, $v_{qq}$, $v_{qs}$ and $v_{ss}$ are the
relative probabilities for different baryon production, they can be
found from quark combinatorics \cite{AnSh,CS}. The contribution of the 
graph in Fig.~4c has a small coefficient $\varepsilon$ and 
$\alpha_{SJ}$ is a parameter of Regge-trajectory for SJ exchange.

The probability to find a comparatively slow SJ in the case of
Fig.~4c can be estimated from the data on the $\bar{p}p$ annihilation
into mesons \cite{KKKY,14r}. This probability is known experimentally
at comparatively small energies only where it is proportional to
$s^{\alpha_{SJ}-1}$ with $\alpha_{SJ} \sim 0.5$. However, it has been
argued \cite{14r} that the annihilation cross section contains a small
piece which is independent of $s$ and thus $\alpha_{SJ}\sim1$.

In \cite{ACKS} the value $\alpha_{SJ}=0.5$ was used. However, for such
value of $\alpha_{SJ}$ different values of $\varepsilon$ were needed for
the description of the experimental data at moderate and high energies.
This problem was solved in \cite{SJ1}, where it was shown with the help
of the new experimental data that all the data can be described with the
parameter values
\begin{equation}
\alpha_{SJ}\, =\, 0.9\,, \quad \varepsilon\, =\, 0.024\,.
\end{equation}

In the situation when two valence quarks of the incident diquark 
fragment into different secondaries (Figs. 4b and 4c) there exist 
mesons $M$ which consist from an initial valence quark and sea 
antiquark. In the case shown in Fig.~4b these exists one such meson and 
in the case of Fig.~4c two such mesons are produced. The fragmentation 
function of a meson $M$ in Fig.~4b was accounted for in all previous 
calculations in the framework of the QGSM \cite{KTM}--\cite{Sh1} (see 
also \cite{22r}). The same fragmentation function can be used to account
for the production of one meson $M$ in Fig.~4c. To account for the 
production of the second meson $M$ in Fig.~4c we need the additional 
fragmentation function which can be written as
\begin{equation}
G^M(z)\ =\ 8 \varepsilon a_0 z^2 (1-z)^2\ .
\end{equation}
The value of the factor $8\varepsilon a_0$ is determined by normalization
condition.

It is necessary to note that the process shown in Fig.~4c can be
realized very naturally in the quark combinatoric approach \cite{AnSh}
with the specified probabilities of a valence quark recombination
(fusion) with sea quarks and antiquarks.

\section{Comparison with the data on nucleon target}

The processes of the baryon charge transfer (Fig.~4c) were not
accounted for in  papers \cite{KTM},
 [3--6]. So first
of all we present the description of data for secondary baryon
production which were described earlier without SJ contribution.
The results for secondary meson and antibaryon production are
the same as before, the accounting for additional mesons $M$
production in the process Fig.~4c gives numerically small
correction.

The inclusive spectra of secondary protons produced in $pp$ collisions
at lab. energies 100 and 175~GeV \cite{Bren} are shown in Fig.~5a
together with the curves calculated in the QGSM. The difference  of
the calculated results with and without SJ contribution is small
here.

\begin{figure}[h]
\centering
\includegraphics[width=.45\hsize]{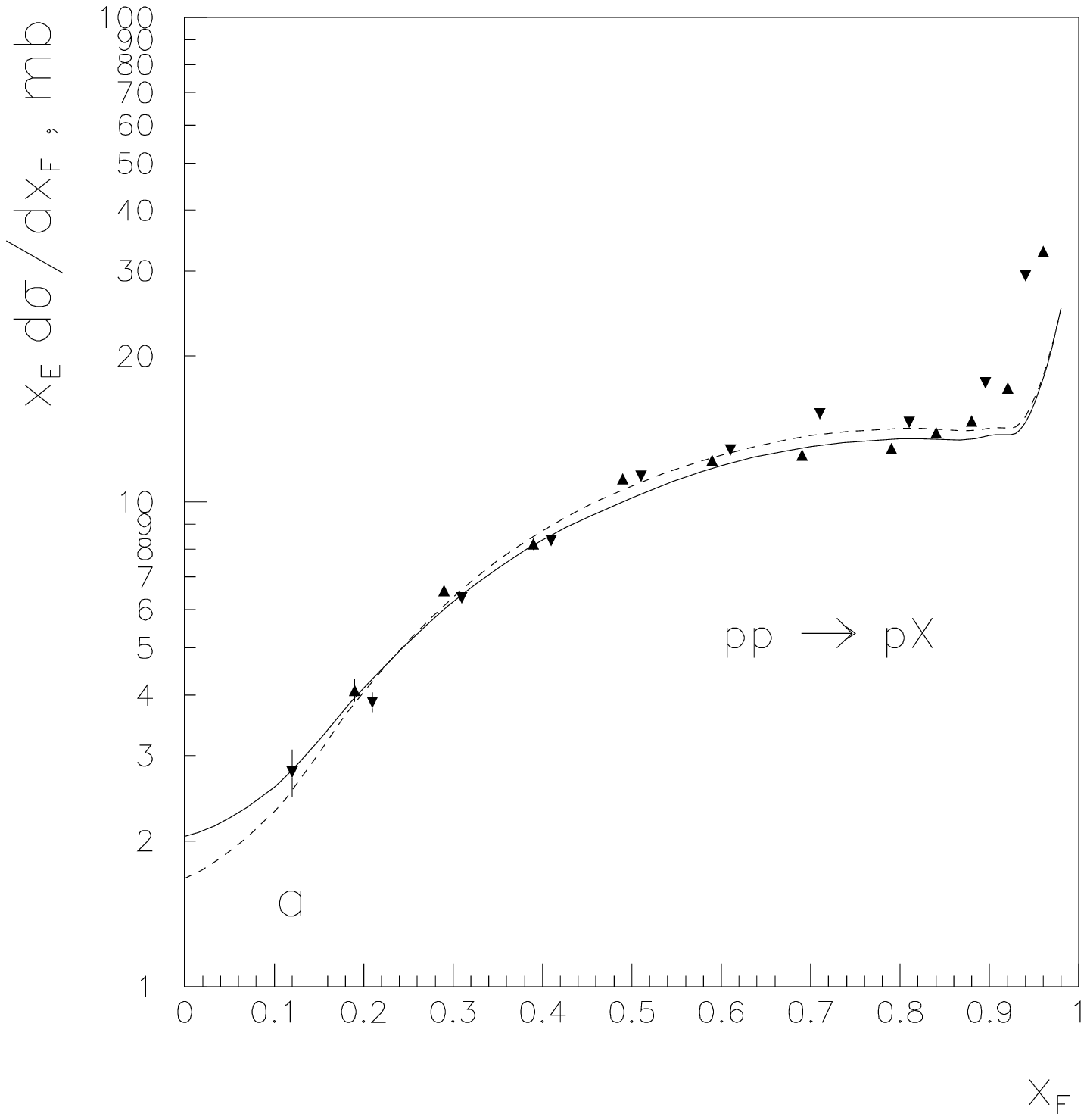}
\includegraphics[width=.45\hsize]{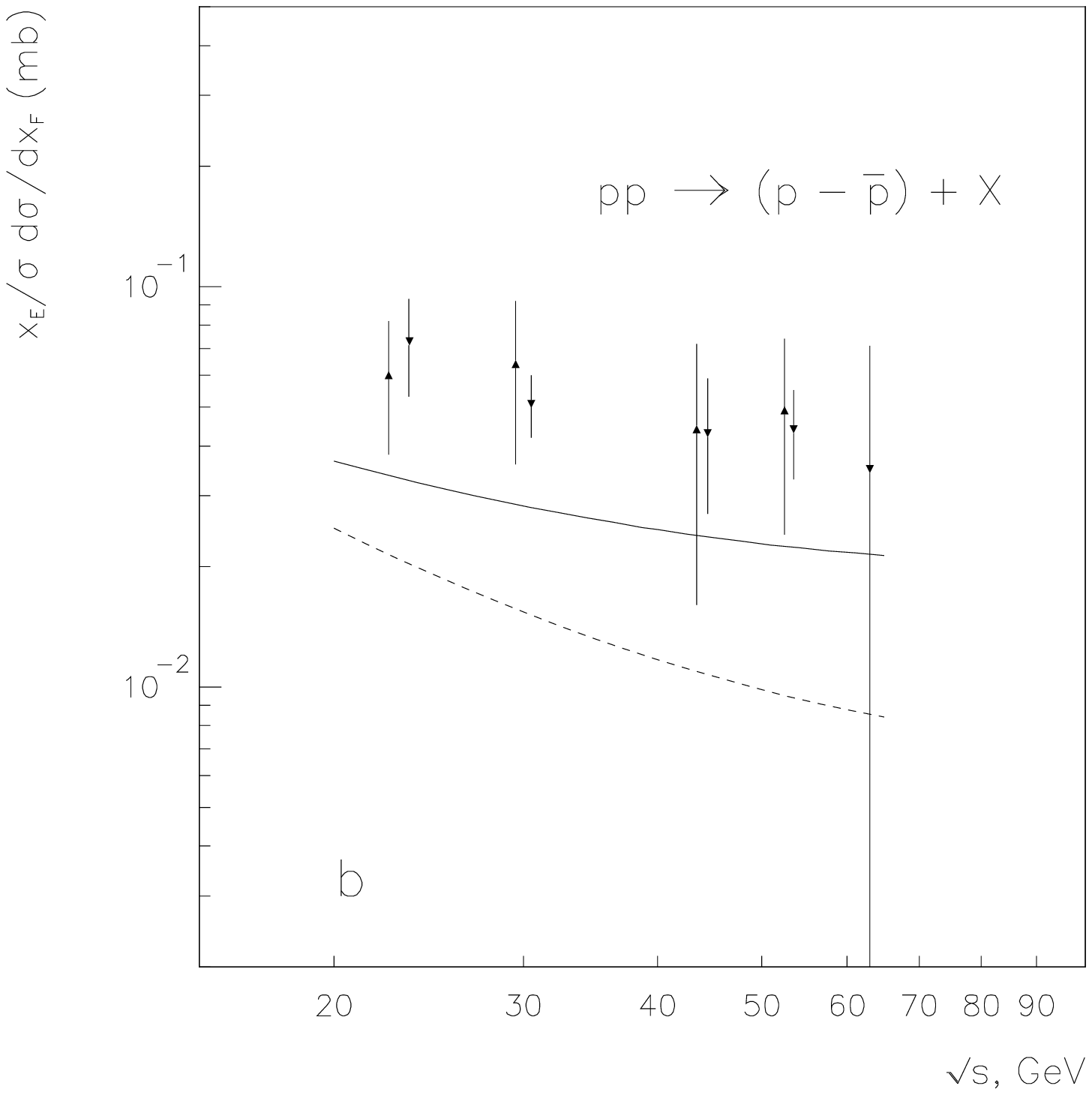}

\vspace*{-0.5cm}
\caption{\footnotesize
Spectra of secondary protons produced in $pp$
collisions at 100 and 175\,GeV/c \cite{Bren} (a). The difference of
the spectra of secondary protons and antiprotons produced in $pp$
collisions at ISR energies \cite{ISR} at $90^o$ in c.m.s. (b). The
QGSM description with $\varepsilon=0.024$ and with
$\varepsilon=0$ are shown by solid and dashed curves,
respectively}
\end{figure}

The data on secondary proton and antiproton production in $pp$
collisions at ISR energies \cite{ISR} at $90^o$ in c.m.s. were
obtained in \cite{ISR}. Their differences, which are more
sensitive to the baryon charge transfer, are presented in Fig.~5b.
Now the effects of SJ contribution are more important. One can see
that the last data are described quite reasonably by QGSM with
$\varepsilon=0.024$, whereas the calculations without SJ
contributions underestimate the data.  However, it is necessary to
note that the systematical errors in \cite{ISR} are of the order of
30\%, and there exists the disagreement between the data \cite{Bren},
\cite{AB} and \cite{ISR} of the order of 20--30\%.

There exist only a few data on secondary production in $pp$ collisions
at RHIC energies. In Fig.~6, we present the rapidity (in c.m.)
distribution of the ratio $\bar{p}/p$ in $pp$ interactions at
$\sqrt{s}=200\,$GeV \cite{BRA}. The QGSM calculation with the SJ
contribution (15) (solid curve) is in reasonable agreement with the
data, and the same calculation without SJ contributions (dashed curve)
overestimates the discussed ratios.

\begin{figure}[h]
\centering
\vskip -0.8cm
\includegraphics[width=.6\hsize]{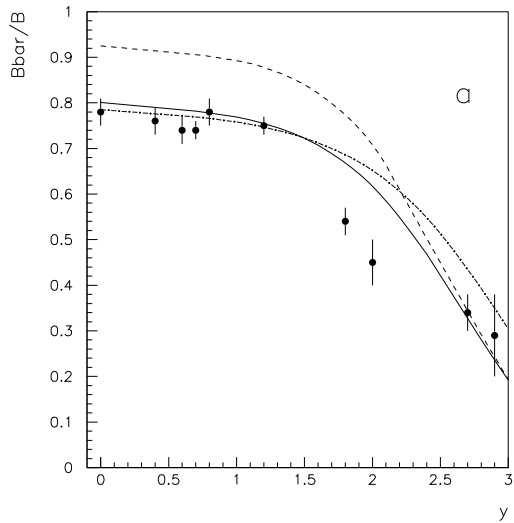}
\vskip -0.8cm
\caption{\footnotesize
Rapidity dependence of $\bar{p}/p$ ratios
for $pp$ collisions at $\sqrt{s}=200\,$GeV. Solid and dashed
curves show the QGSM description with and without SJ contribution
and dash-dotted curve shows the QGSM predictions for
$\bar{\Lambda}/\Lambda$ ratio}
\vskip -0.3cm
\end{figure}

It is important to note that at asymptotic energies the ratio 
$\bar{p}/p$ is equal to the unity. So the deviation of the discussed 
ratio from the unity can appear due to some physical reason. One
can see from Fig.~6 that the SJ contribution changes the deviation of
$\bar{p}p$ from unity at small $y$, {\em i.e.} in the central region
about three times in comparison with the calculation without SJ
contribution.

The dash-dotted curve in Fig.~6 shows the QGSM with SJ contribution
predictions for $\bar{\Lambda}/\Lambda$ ratios at RHIC energies. The
spectra of secondary $\Lambda$ produced in $pp$ collisions at fixed
target energies depend rather weakly on the SJ diffusion effects, as 
one can see in \cite{ACKS}. The QGSM calculations \cite{SJ1} predict
practically equal values of $\bar{B}/B$ ratios in midrapidity region
for all strange baryons that is qualitatively confirmed by the RHIC
data on $Au-Au$ collisions \cite{MuNa}.

The data of baryon production in the pion fragmentation region are
more sensitive to the SJ contribution in comparison with $pp$
collisions. Significant effects come from the possibility of the baryon
number to transfer from the target nucleon to the beam fragmentation
region. The existing data for secondary antinucleon production 
\cite{Bren} are presented in Fig.~7. The spectra of antiprotons produced 
in $\pi^-p$ collisions shown in Fig.~7a allow one to fix the 
fragmentation function of quark into baryon/antibaryon. Were the 
contribution of the baryon charge transfer negligibly small, the 
inclusive spectra of reactions $\pi^-p\to\bar{p}X$ and $\pi^+p\to pX$ in 
the pion fragmentation region would be practically the same \cite{Sh}. 
Actually, the data for the second reaction are significantly higher than 
for the first one providing evidence for the baryon charge transfer due 
to the SJ diffusion. The difference of the inclusive spectra in the two 
considered processes allows one to estimate quantitatively the 
contribution of the baryon charge transfer. The QGSM calculations with 
accounting for SJ contribution are in better agreement with the data.
The experimental data for secondary $\Lambda$ and $\bar\Lambda$
production in $\pi p$ and $Kp$ collisions also are in agreement
\cite{AMS,AMS1} with the discussed approach predictions, but the
experimental error bars are rather large.

\begin{figure}[htb]
\centering
\includegraphics[width=.45\hsize]{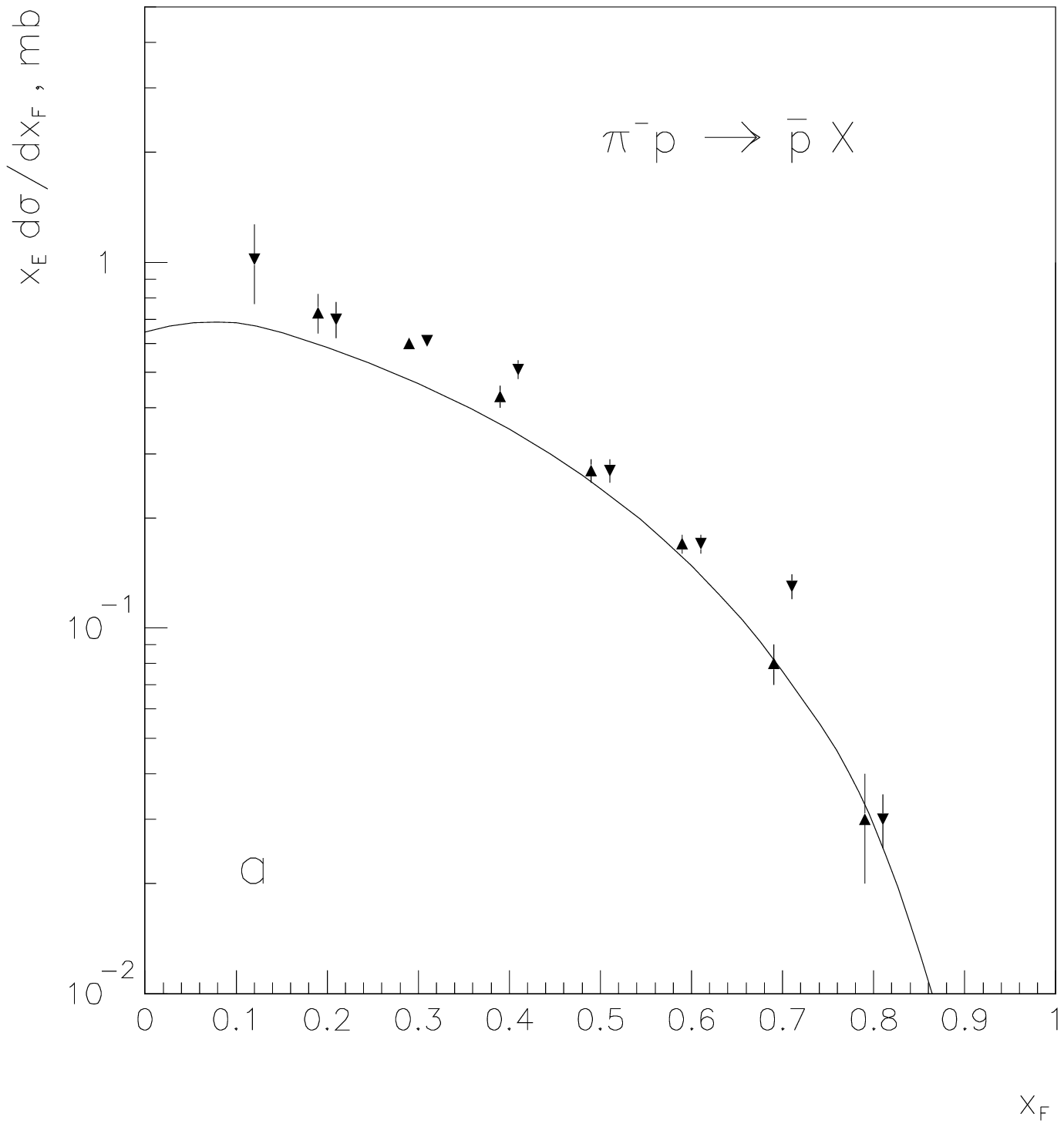}
\includegraphics[width=.45\hsize]{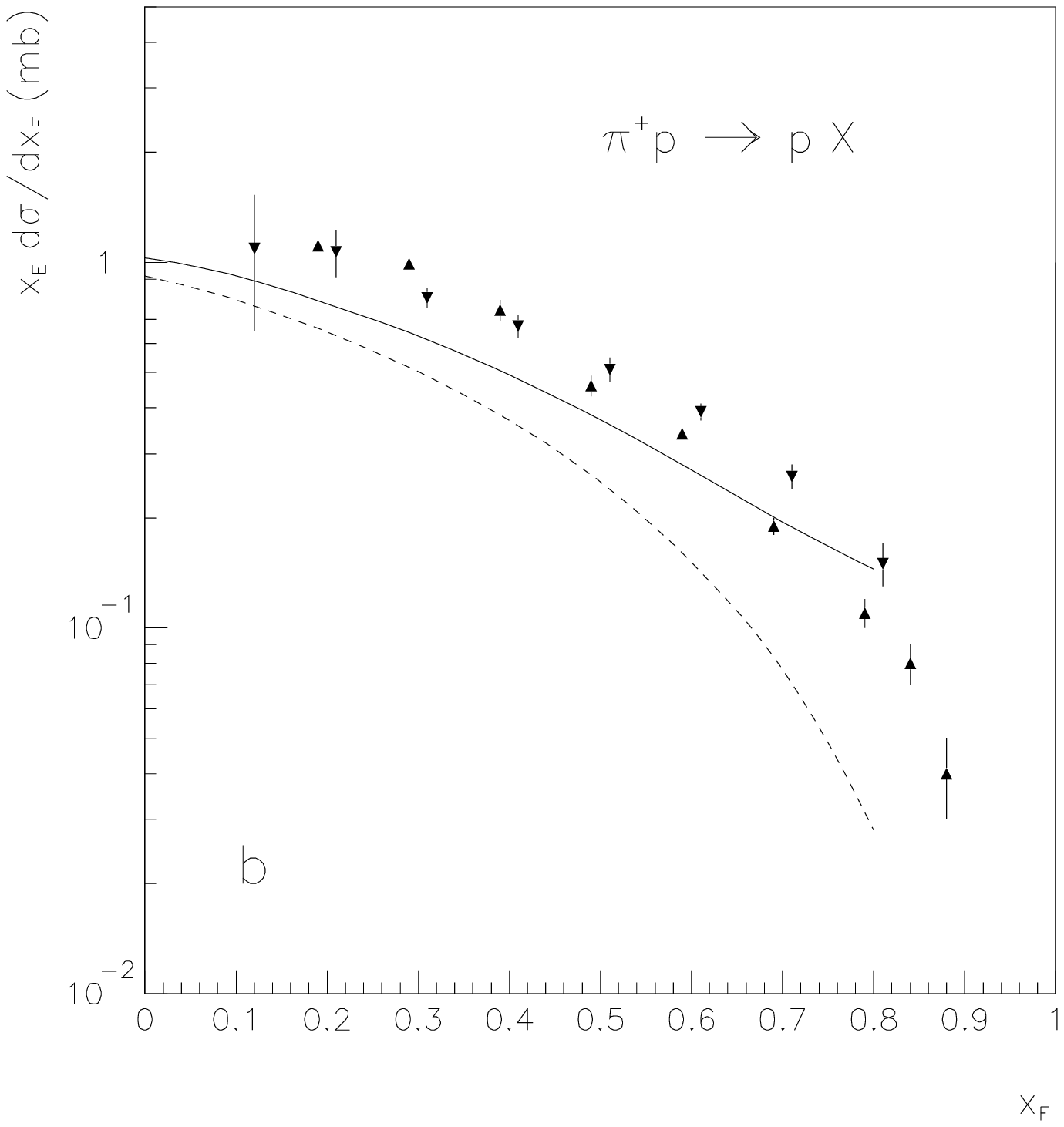}

\vspace*{-0.5cm}
\caption{\footnotesize
The spectra of secondary antiprotons in $\pi^-p$ collisions
(a) and of protons in $\pi^+p$ collisions (b) at lab. energies 100
and 175\,GeV \cite {Bren}. The QGSM description with
$\varepsilon=0.024$ and with $\varepsilon=0$ in (b) is shown by solid
curve and by dashed curve, respectively}
\end{figure}
\begin{figure}[htb]
\centering
\includegraphics[width=.35\hsize]{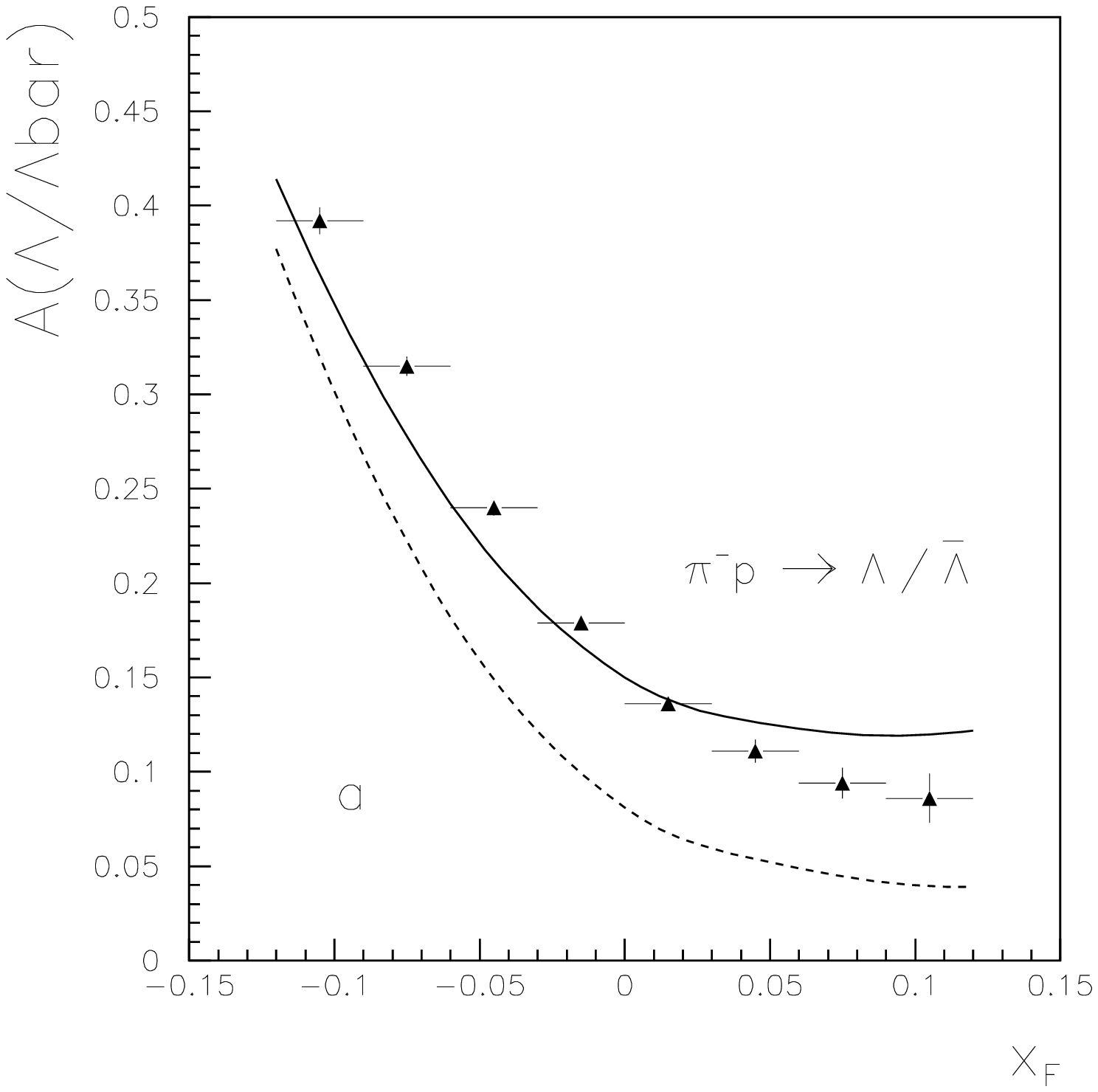}
\includegraphics[width=.35\hsize]{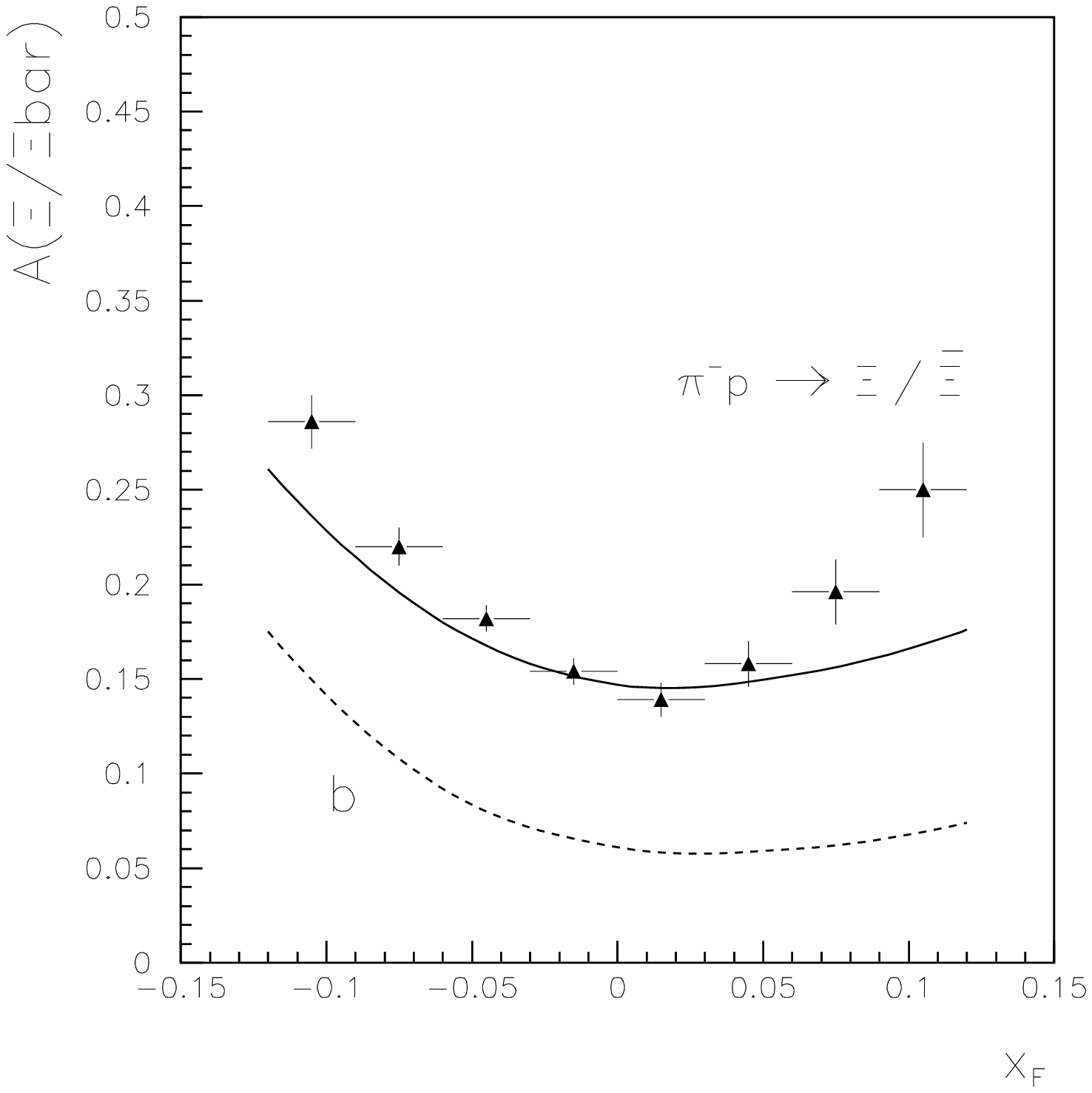}
\includegraphics[width=.35\hsize]{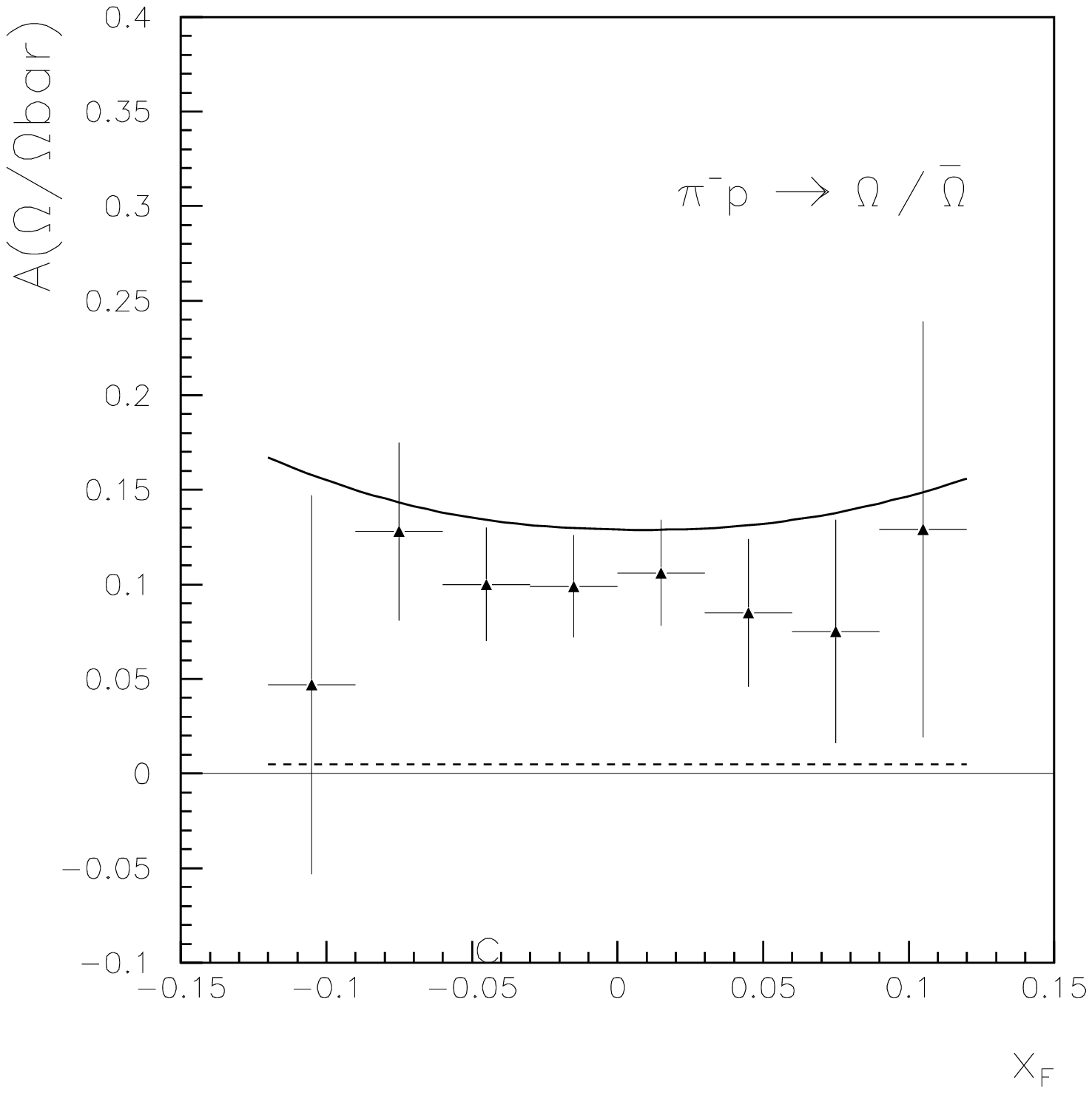}

\vspace*{-0.3cm}
\caption{\footnotesize
The asymmetries of secondary $\Lambda/\bar{\Lambda}$ (a),
$\Xi^-/\overline{\Xi}^+$ (b), and $\Omega/\bar{\Omega}$ (c), in $\pi^-p$
collisions at 500\,GeV/c \cite{ait} and its description by QGSM model.
For all cases the calculations with $\epsilon$ = 0.024 are shown by solid
curves and the variants with $\epsilon=0$ are shown by dashed curves}
\end{figure}

In Fig.~8, we show the data \cite{ait} on the asymmetry of strange
baryons produced in $\pi^-$ interactions\,\footnote{These data were
obtained from pion interactions on a nuclear target where
different materials were used in a very complicated geometry. We
assume that the nuclear effects are small in the asymmetry and compare
the pion-nucleus data with calculations for $\pi^-p$ collisions.} at
500\,GeV/c. The asymmetry is determined as
\begin{equation}
A(B/\bar{B})\ =\ \frac{N_B-N_{\bar{B}}}{N_B+N_{\bar{B}}}
\end{equation}
for each $x_F$ bin.

Theoretical curves for the data on all asymmetries calculated
with $\varepsilon=0.024$ are in reasonable agreement with the data.
In case of $\Omega/\bar{\Omega}$ production we predict, following
\cite{Noda}, a non-zero asymmetry in agreement with experimental
data. Let us note that the last asymmetry is absent, say, in the naive
quark model or in recombination model because $\Omega$ and
$\bar{\Omega}$ have no common valence quarks with the incident
particles.

Preliminary data on $p/\bar{p}$ asymmetry in $ep$ collisions at
HERA were presented by the H1 Collaboration \cite{H1}. Here the
asymmetry is defined as
\begin{equation}
A_B\ =\ 2 \frac{N_p-N_{\bar{p}}}{N_p + N_{\bar{p}}}\, ,
\end{equation}
{\em i.e.} with an additional factor 2 in comparison with Eq. (17). The
experimental value of $A_B$ is equal to $(8.0\pm1.0\pm2.5)\%$
\cite{H1} for secondary baryons produced at $x_F\sim0.04$ in the
$\gamma p$ c.m. frame. QGSM without SJ contribution, {\em i.e.} with
$\varepsilon=0$ predicts here only 2.9\,\%, which is significantly
smaller than the experimental value, whereas the calculation with
$\varepsilon = 0.024$ gives the value 9.9\,\%, in good agreement with
data.

\section{Production of secondaries from nuclear \newline targets in QGSM}

As was mentioned above, the high energy hadron--nucleon and 
hadron--nucleus interactions are considered in the QGSM and in DPM as 
proceeding via the exchange of one or several Pomerons. Each Pomeron 
corresponds to a cylindrical diagram, and thus, when cutting a Pomeron, 
two showers of secondaries are produced. The inclusive spectrum of 
secondaries is determined by the convolution of diquark, valence quark 
and sea quark distributions $u(x,n)$ in the incident particles and the 
fragmentation functions $G(z)$ of quarks and diquarks into secondary 
hadrons.

The diquark and quark distribution functions depend on the number $n$ of
cut Pomerons in the considered diagram. In what follows we use the
formalism of QGSM described in Section~3.

In the case of nuclear targets we should consider the possibility of
one or several Pomeron cuts in each of the $\nu$ blobs of
hadron--nucleon inelastic interactions as well as cuts between Pomerons.
For example, for the $\pi A$ collision one of the cut Pomerons links 
with valence antiquark and the valence quark of the projectile pion with
valence quark and diquark of one target nucleons. The other Pomerons 
link with the sea quark--antiquark pairs of the projective pion with 
diquarks and valence quarks of another target nucleons and with sea 
quark--antiquark pairs of the target.

For example, one of the diagram for inelastic interaction with two
target nucleons is shown in Fig.~9. In the blob of the $\pi N_1$
inelastic interaction one Pomeron is cut, and in the blob of $\pi
N_2$ interaction two Pomerons are cut. It is essential to take
into account every possible Pomeron configuration and permutation
for all diagrams. The process shown in Fig.~9 satisfies the
condition \cite{Sh2} that the absorptive parts of hadron--nucleus
amplitude are determined by the combinations of the absorptive parts of
hadron--nucleon interactions.

\begin{figure}[h]
\centering
\includegraphics[width=.5\hsize]{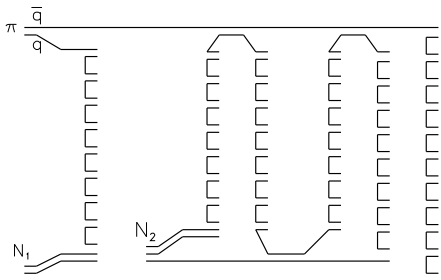}

\vspace*{-0.3cm}
\caption{\footnotesize
One of the diagrams for inelastic interaction of incident pion with 
two target nucleons $N_1$ and $N_2$ in  $\pi A$ collision}
\end{figure}

In the case of inelastic interactions with $\nu$ target nucleons, let
$n$ be the total number of cut Pomerons in $hA$ collisions ($n\ge\nu$)
and let $n_i$ be the number of cut Pomerons connecting with the $i$-th
target nucleon ($1\le n_i\le n-\nu+1$). We define the relative
weight of the contribution with $n_i$ cut Pomerons in every $hN$ blob
as $w^{hN}_{n_i}$. For the inclusive spectrum of the secondary hadron
$h$ produced in a $\pi A$ collision we obtain \cite{KTMS}
\begin{eqnarray}
\frac{x_E}{\sigma^{prod}_{\pi A}} \frac{d \sigma}{dx_F} & = &
\sum^A_{\nu=1} V^{(\nu)}_{\pi A} \left\{ \sum^{\infty}_{n=\nu}
\sum^{n-\nu+1}_{n_1 = 1} \cdots
\sum^{n-\nu+1}_{n_{\nu}=1} \prod^{\nu}_{l=1} w^{\pi N}_{n_l} \right.
 \nonumber\\
& \times & \bigg[f^h_{\bar{q}}(x_+,n)f^h_q(x_-,n_l) +
f^h_q(x_+,n)f^h_{qq}(x_-,n_l)
\nonumber\\
& + & \left. \sum^{2n-2}_{m=1} f^h_s(x_+,n)f^h_{qq,q,s}(x_-,n_m)\bigg]
 \right\},
\end{eqnarray}
where $V^{(\nu)}_{pA}$ is the probability of ``pure inelastic"
(nondiffractive) interactions with $\nu$ target nucleons, and we should
account for all possible Pomeron permutation and the difference in quark
content of the protons and neutrons in the target.

In particular, the contribution of the diagram in Fig.~9 to the
inclusive spectrum is

\newpage

\begin{eqnarray}
\frac{x_E}{\sigma^{prod}_{\pi A}}
\frac{d\sigma}{dx_F} &=& 2 V^{(2)}_{\pi A} w^{\pi N_1}_1w^{\pi N_2}_2
\bigg\{ f^h_{qq}(x_+,3)f^h_q(x_-,1)
 + f^h_q(x_+,3)f^h_{qq}(x_-,1)
\nonumber\\
&+& f^h_s(x_+,3)\left[f^h_{qq}(x_-,2) +f^h_q(x_-,2)
+ 2f^h_s(x_-,2)\right]  \bigg\}.
\end{eqnarray}

In the case of a nucleon beam the valence antiquark contributions of
incident particle should be substituted by the contribution of valence
diquarks.

The diquark and quark distributions as well as the fragmentation
functions are here the same as in the case of the nucleon target (see
Section 3). We account all three possibilities that the secondary baryon
can consist of the SJ together with two valence and one sea quarks
(Fig.~4a), with one valence and two sea quarks (Fig.~4b) or with
three sea quarks (Fig.~4c).

In the present calculations, following the present experimental data,
we increase the portion of strange quarks in the sea, $S/L$
\cite{ACKS,CS} from $S/L = 0.2$ to $S/L = 0.32$.
\begin{figure}[h]
\vskip -0.5cm
\centering
\includegraphics[width=.5\hsize]{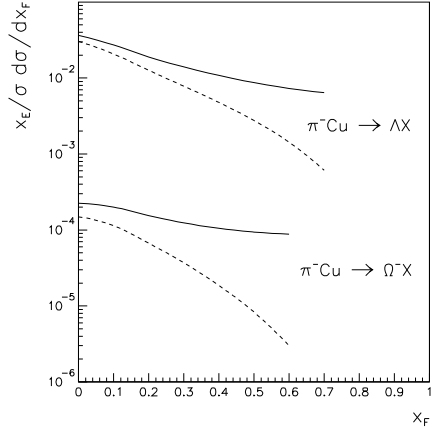}
\vskip -0.5cm
\caption{\footnotesize
The QGSM predictions for the inclusive cross sections of
$\Lambda$ and $\Omega^-$ production in $\pi^-$Cu collisions
at 400\,GeV/c with (solid curves) and without (dashed curves) SJ
contributions}
\end{figure}

To illustrate the expected effects of SJ contributions we present in
Fig.~10 the predicted inclusive cross sections of
$\pi^-{\rm Cu}\to\Lambda X$ and $\pi^-{\rm Cu}\to\Omega^-X$ reactions 
with (solid curves) and without (dashed curves) SJ contributions shown 
in Fig.~4c. We discuss precisely these reactions because the secondary
baryons and the correspondent antibaryons $\bar\Lambda$ and
$\overline{\Omega}^+$ have symmetrical quark states in respect to the
incident $\pi^-$. So the SJ contribution, which is equal to the 
difference between solid and dashed curves in Fig.~10, can be measured
experimentally as the difference in $\Lambda-\bar\Lambda$, or
$\Omega^- -\overline\Omega^+$ production at high energies.

In general, the SJ contribution shown in Fig.~4c increases the inclusive
cross sections of $\Lambda$ and $\Omega^-$ production. The spectra of
antibaryons are not affected. However, numerically these effects are
rather small, for example, mean multiplicity of secondary $\Lambda$ in
forward hemisphere should increase in about 15\% that should be 
compensated by the correspondent decrease of secondary nucleon 
multiplicity in the target fragmentation region.

\vspace*{-0.5cm}

\section{Comparison with the nuclear target data}

In Fig.~11, we show the data \cite{WA97} on the midrapidity inclusive
densities, $dn/dy$ at $\vert y_{c.m.} \vert < 0.5$ of secondary
$\Lambda$, $\bar\Lambda$, $\Xi^-$, $\overline{\Xi}^+$ and the sum
$\Omega^- +\overline{\Omega}^+$ produced in $p$Be and $p$Pb collisions
at 158\,GeV/c. These data are in reasonable agreement with the QGSM
calculations \cite{SJ2} and this agreement is better with the account 
for the SJ contributions for secondary baryons.

\begin{figure}[h]
\centering
\includegraphics[width=.5\hsize]{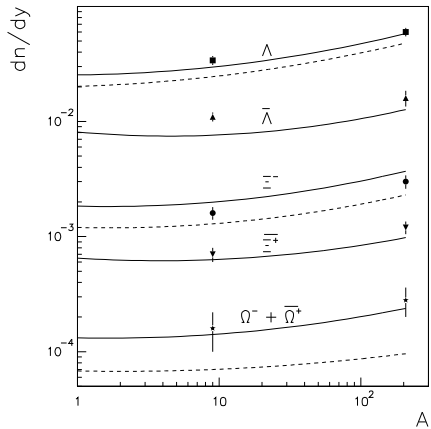}
\vskip-0.5cm
\caption{\footnotesize
Yields of $\Lambda$ (closed squares), $\bar\Lambda$ (triangles), 
$\Xi^-$ (points), $\overline{\Xi}^+$ (turned over triangles) and the 
sum $\Omega^- + \overline{\Omega}^+$ (stars) per unit of rapidity at 
central rapidity as a function of the target atomic weight for $pA$ 
collisions at 158\,GeV/c. The QGSM predictions with SJ contribution are 
shown by solid curves and without SJ contribution by dashed curves}
\end{figure}

Similarly to the case of a nucleon target, the SJ effects are more
important in the meson beam fragmentation. In Fig.~12 we present the
NA49 Coll. data \cite{NA49} on the $x_F$ distributions of net protons
$(p-\bar{p})$ produced in $\pi p$ and $\pi$Pb interactions at
$\sqrt{s}$ = 17.2\,GeV. The beam $\pi $ is determined in \cite{NA49} as
$(\pi^+ +\pi^-)/2$. The data are described rather good \cite{SJ2} with 
the account for the SJ diffusion (solid curves in Fig.~12) and the 
variant without SJ contribution (dashed curves) underestimates the data 
in several times at $x_F>0.1$.

\begin{figure}[h]
\centering
\includegraphics[width=.55\hsize]{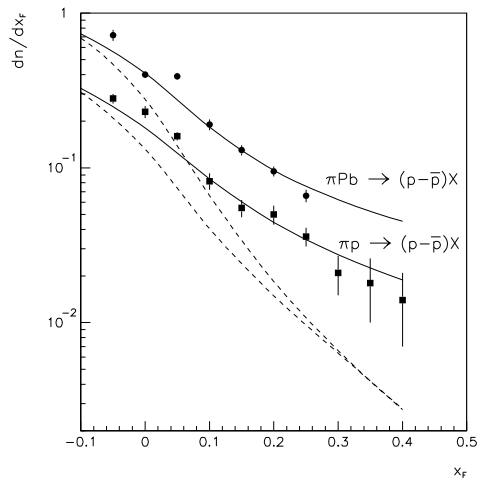}
\vskip -0.5cm
\caption{\footnotesize
Feynman-$x$ distributions of net protons produced in $\pi p$
(squares) and $\pi$Pb (points) interactions at $\sqrt s=17.2\,$GeV.
Solid and dashed curves show the QGSM description with and without SJ
contribution, respectively}
\end{figure}

The experimental data of WA89 Coll. \cite{WA89} on $\Lambda$, $\Xi^-$,
$\bar\Lambda$ and $\overline{\Xi}^+$ production from C and Cu
targets by 345\,GeV/c $\pi^-$ beam are shown in Fig.~13. The yields of
secondary hyperons are in reasonable agreement with QGSM predictions
\cite{SJ2} accounting for the SJ contributions (solid curves in 
Figs. 13a and 13b). The calculations without SJ contributions (dashed 
curves) disagree with the data.

\begin{figure}[h]
\centering
\includegraphics[width=.36\hsize]{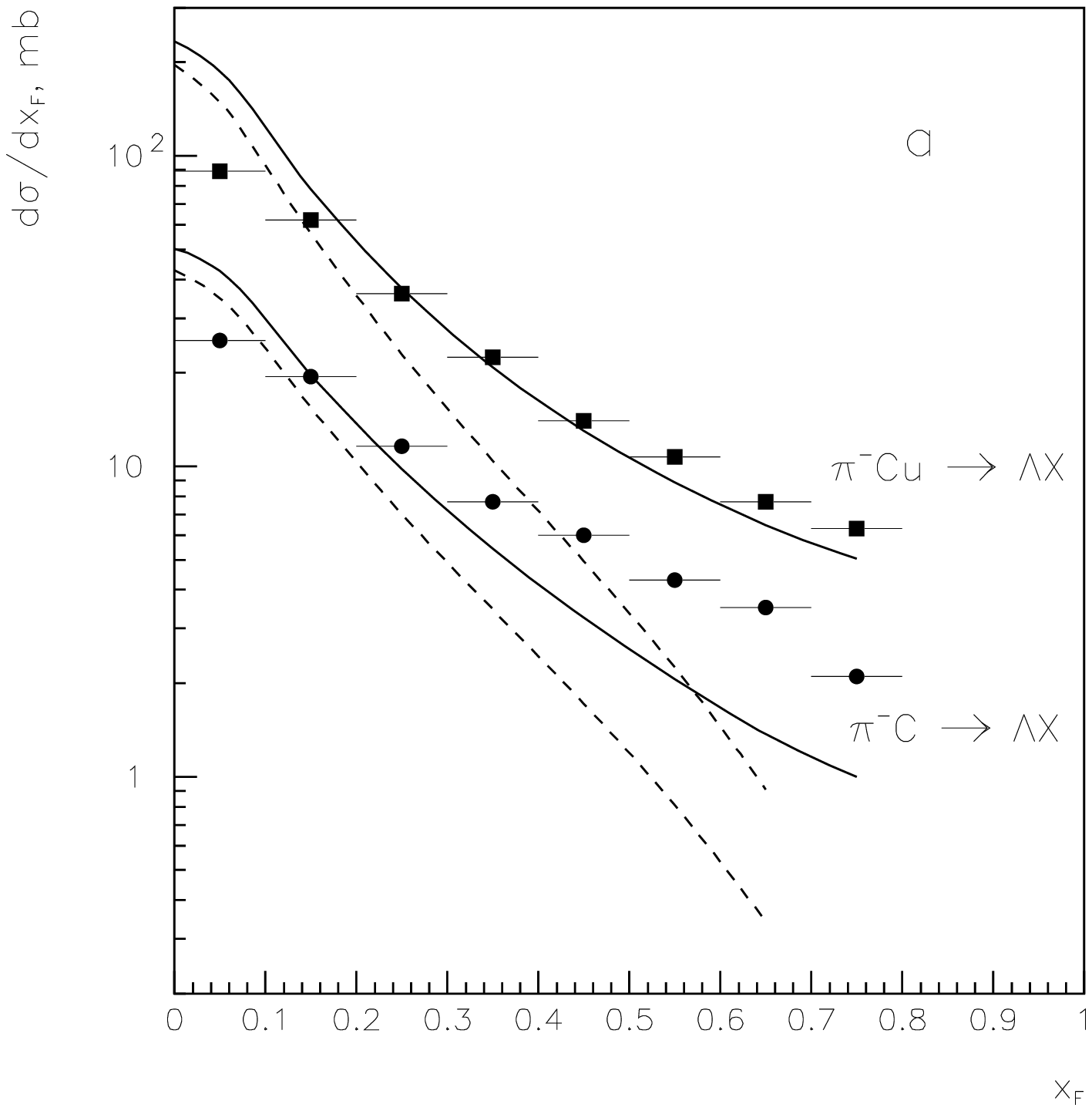} 
\includegraphics[width=.36\hsize]{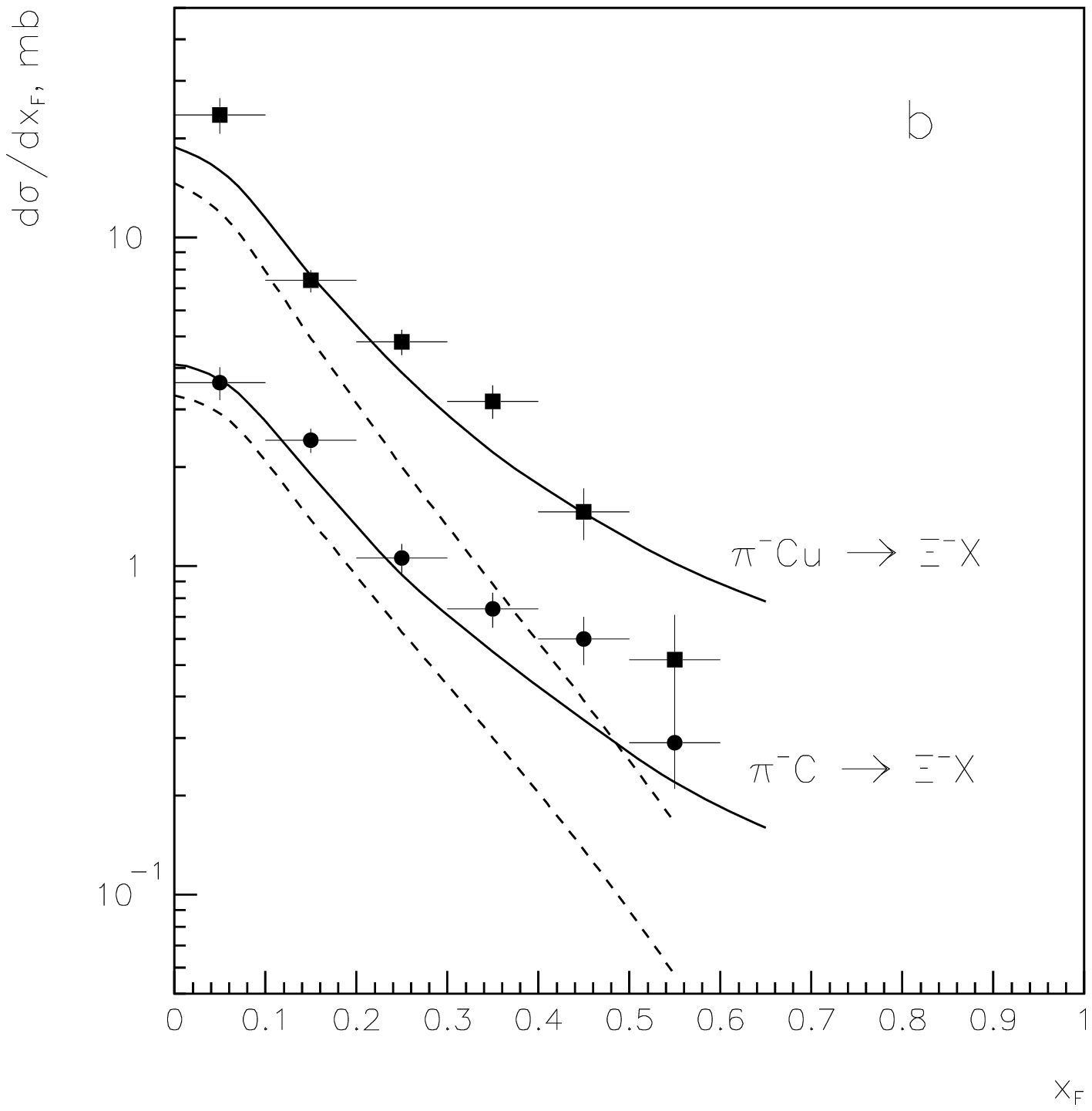} 
\includegraphics[width=.36\hsize]{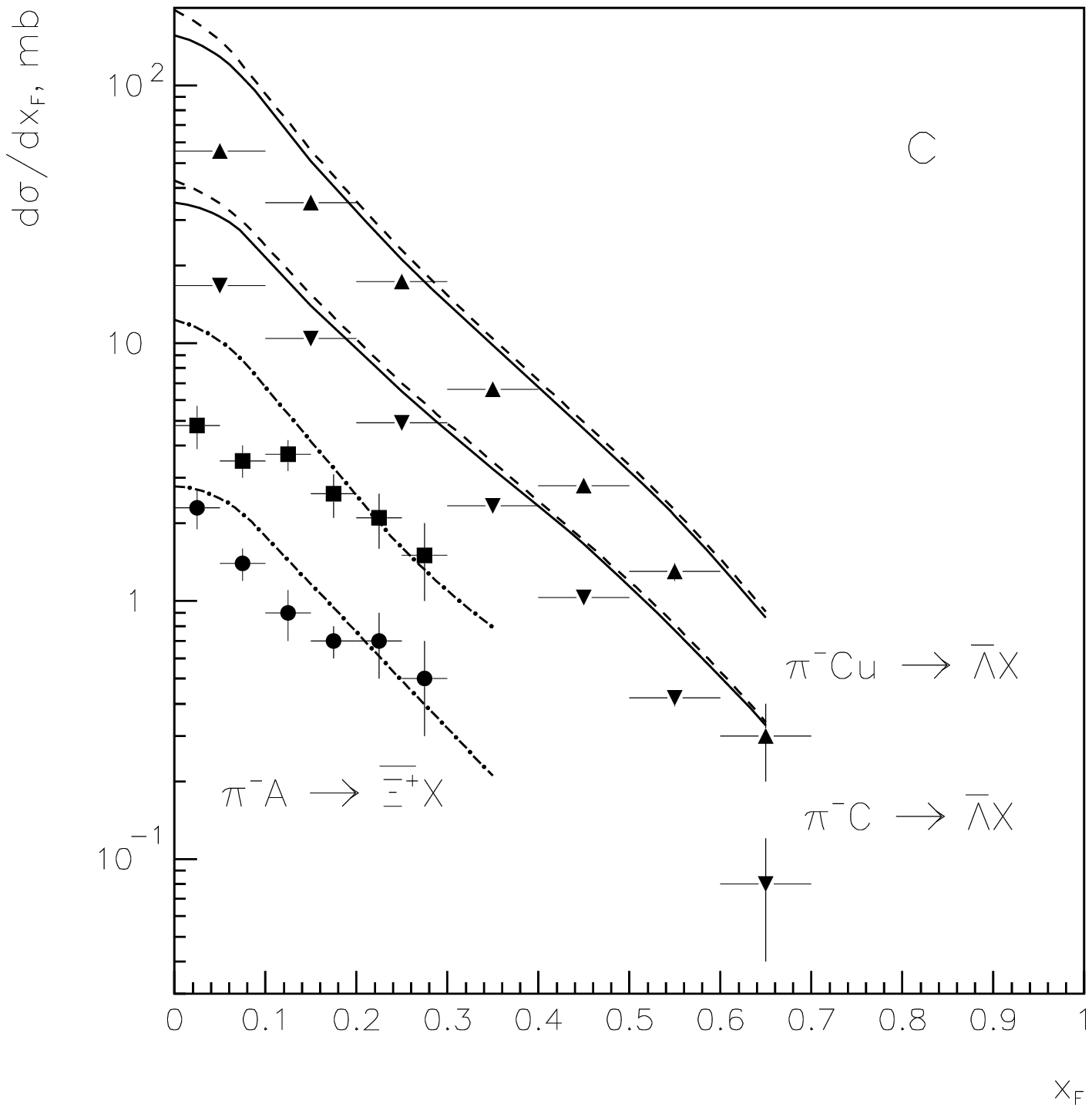} 
\caption{\footnotesize
Feynman-$x$ distributions of secondary $\Lambda$ (a), $\Xi^-$
(b), $\bar{\Lambda}$ and $\overline{\Xi^+}$ (c) produced in $\pi^-$C
and $\pi^-$Cu interactions at 345\,GeV/c. Solid and dashed curves show
the QGSM prediction for secondary hyperon spectra with and without SJ
contribution}
\end{figure}

The yields of $\bar{\Lambda}$ and $\overline{\Xi}^+$ \cite{WA89}, which
do not depend on SJ contribution, are shown in Fig.~13c. These data are
described by QGSM on the reasonable level.

The data presented in \cite{WA89} allow one to calculate the asymmetries
of secondary $\Lambda/\bar\Lambda$ production defined by Eq.~(17).
They are presented in Fig.~14a for the cases of $\pi^-$Cu (points)
and $\pi^-$C (squares) interactions. The curves show the QGSM
calculations \cite{SJ2} for copper (dotted curve), carbon (dashed curve) 
and nucleon (solid curve) targets. We predict some $A$-dependence
of the asymmetry for beam fragmentation region. The agreement with the
data is reasonable in the central region, but we obtain some
underestimation of asymmetry for $x_F>0.3$.

In Fig.~14b we present the data of \cite{E769} for the same asymmetry
Eq.~(17) obtained for $\pi^-$ interactions with multifoil target with
different atomic weights, see \cite{E769}. Here the QGSM predictions
\cite{ACKS,SJ2} even for $\pi^-p$ interactions (solid curve), {\em i.e.} 
neglecting the $A$-dependence, overestimate the data at $x_F>0.1$. In 
the central region, $\vert x_F\vert\le0.1$, our calculations agree with 
the data of both \cite{WA89} and \cite{E769} as well as of \cite{ait}. 
Here we predict the practical absence of $A$-dependence (or weak 
dependence) for $\Lambda/\bar\Lambda$ asymmetries, as it was assumed in 
\cite{ACKS}. In the $\pi^-$ fragmentation region the data of \cite{WA89} 
and \cite{E769} are in strong disagreement with each other.

\begin{figure}[h]
\centering
\includegraphics[width=.45\hsize]{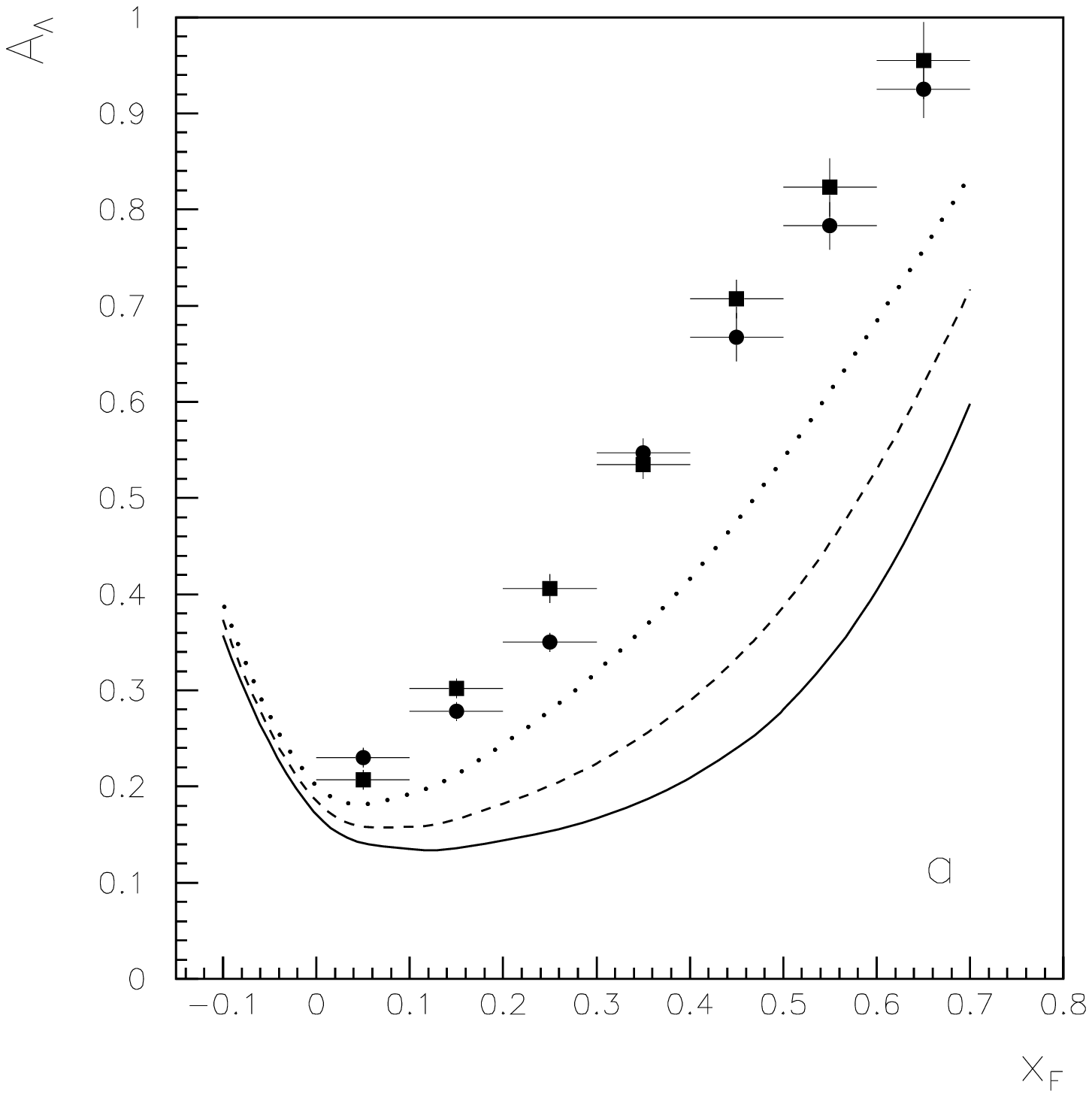}
\includegraphics[width=.45\hsize]{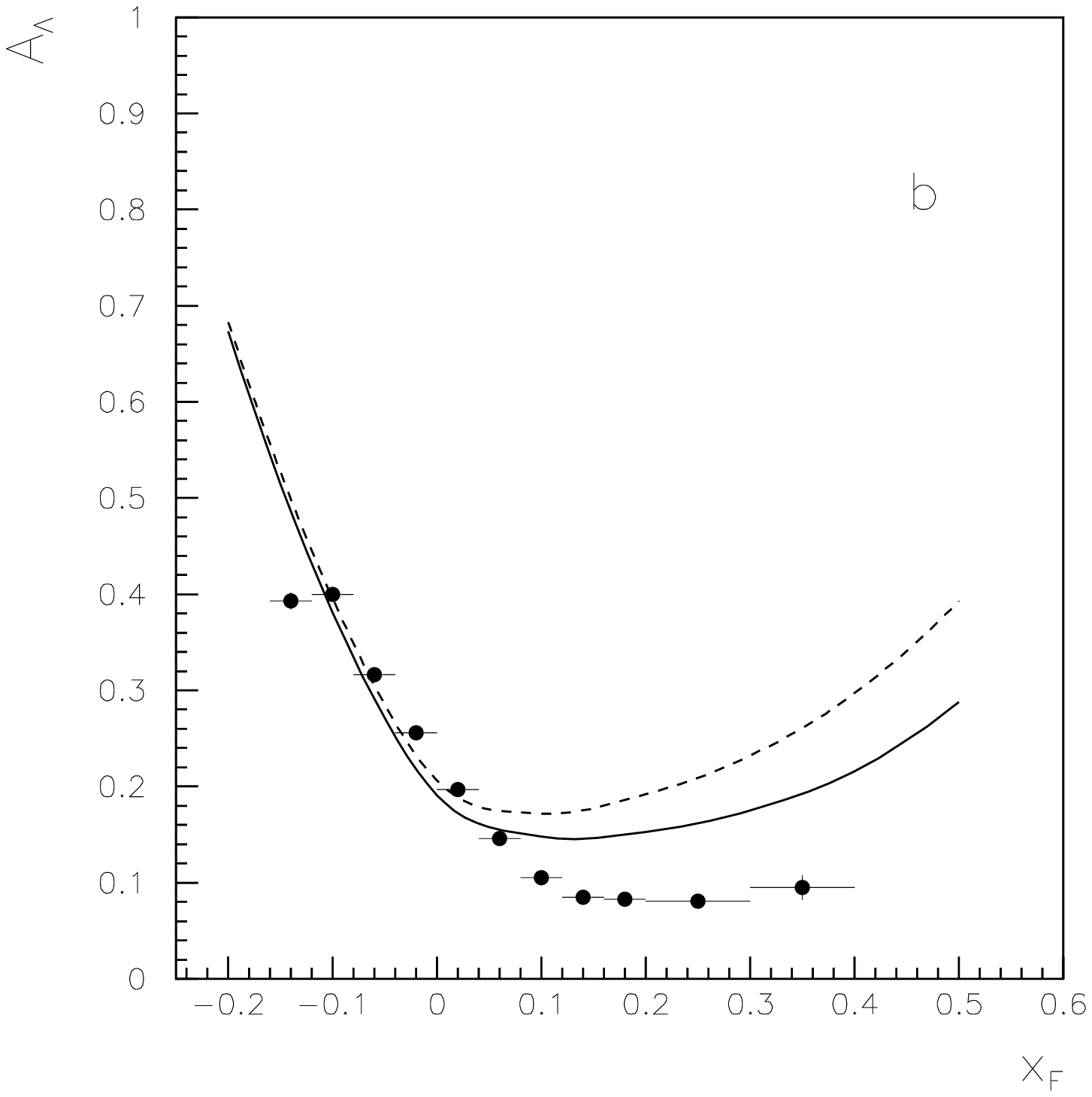}
\vskip -.5cm
\caption{\footnotesize
The asymmetries of secondary $\Lambda/\bar{\Lambda}$
production in $\pi^-$C (squares) and $\pi^-$Cu (points) interactions
at 345\,GeV/c (a). The same asymmetries for $\pi^-A$ collisions at
250\,GeV/c (b). Solid, dashed and dotted curves show the QGSM
predictions for nucleon, carbon and copper targets, respectively}
\end{figure}

The comparison of data shown in Figs. 13a and 13c allows us to
obtain the direct results for SJ contribution to hyperon production
cross section. Really, $\Lambda$ has the valence quark content $uds$,
so the fast incident $\pi^-$ $(\bar{u}d)$ should fragment into
secondary $\Lambda$ and $\bar{\Lambda}$ with equal probabilities,
{\em i.e.}  the $\pi^-\to\Lambda,\bar{\Lambda}$ fragmentation is
flavour symmetrical, contrary, say, to the $\pi^-\to p, \bar{p}$ 
fragmentation.

So the contributions of the processes of Figs. 4a and 4b are negligible
at $x_F>0.1$ and the difference in the spectra of secondary $\Lambda$
and $\bar\Lambda$ determines the SJ contribution of the process shown
in Fig.~4c. This difference is obtained to be rather large in
\cite{WA89} (in the case of nuclear targets) but very small in \cite{Mik}
(for $\pi^-p$ collisions). To show the disagreement between the data of
\cite{WA89} and \cite{Mik} we present in the Table~1 the values of the
parameter $n$ for the parametrization
\begin{equation}
d\sigma/dx_F\ =\ C (1-x_F)^n,
\end{equation}
which were obtained in \cite{WA89} and \cite{Mik} for secondary $\Lambda$
and $\bar{\Lambda}$ production.

\begin{table}
\caption{
The values of the parameter $n$ in Eq.~(21) obtained in \cite{Mik} 
and \cite{WA89} for $\Lambda$ and $\bar{\Lambda}$ production in high
energy $\pi^-p$ and $\pi^-A$ collisions}

\begin{center}
\begin{tabular}{|c|c|}\hline
 Reaction & $n$  \\   \hline
$\pi^-p\to\Lambda$  \cite{Mik} & $2.0 \pm 0.1$\\
$\rm\pi^-C\to\Lambda$  \cite{WA89} & $2.12 \pm 0.02$\\
$\rm\pi^-Cu\to\Lambda$ \cite{WA89} & $2.71 \pm 0.02$   \\
$\pi^-p \to \bar\Lambda$  \cite{Mik} & $2.0 \pm 0.1$  \\
$\rm\pi^-C\to\bar{\Lambda}$  \cite{WA89} & $5.23 \pm 0.04$  \\
$\rm\pi^-Cu\to\bar{\Lambda}$  \cite{WA89} & $5.53 \pm 0.04$  \\
\hline
\end{tabular}
\end{center}
\end{table}

The values of $n$ for the secondary $\Lambda$ production obtained both
in \cite{WA89} and \cite{Mik} on nucleon and on nuclear targets are in
agreement, with a natural weak $A$-dependence. The value of $n$
slightly increases with $A$ that demonstrate well-known effect of
nuclear absorption \cite{KTMS},\cite{Sh2}--\cite{ASS}.
The values of $n$ for
$\bar{\Lambda}$ production obtained in \cite{Mik} and \cite{WA89} are
absolutely different. The data of \cite{Mik} show the absence, or very
small contribution of SJ diffusion in the case of $\Lambda$ and
$\bar{\Lambda}$ production, in contradiction with \cite{WA89} and
with several another results, see, for example, \cite{ait,AMS}.

It is possible to extract the SJ contribution from the experimental
data of \cite{WA89}. At positive $x_F$ the condition $x_F>0.1$ for
$\pi^-$ beam at 345\,GeV/c means $y-y_{\rm target}>4$ for secondary
$\Lambda$, {\em i.e.} these $\Lambda$ are rather far from the target
nucleons in rapidity space. So the difference between the yields of
secondary $\Lambda$ and $\bar{\Lambda}$ comes from the SJ contribution
shown in Fig.~4c. The SJ contributions to the spectrum of secondary
$\Lambda$ in $\pi^-$Cu and $\pi^-$C collisions, obtained by such a way
are presented in Fig.~15.

\begin{figure}[h]
\centering
\vskip -1cm
\includegraphics[width=.55\hsize]{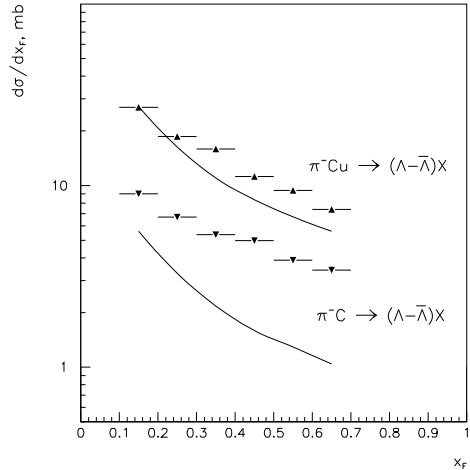}
\vskip -0.5cm
\caption{\footnotesize
The extracted SJ contributions to the spectra of $\Lambda$
in $\pi^-A$ collisions at \newline 345 GeV/c and their description by
QGSM}
\end{figure}

The $x_F$-distributions of the $\Lambda$ produced from copper
target are in reasonable agreement with QGSM calculations, however
in the case of carbon target we obtain the disagreement coming
mainly from not good description in Fig.~13a and some
overestimation of $\bar{\Lambda}$ production in Fig.~13c. It is
necessary to note that the data \cite{Mik} leads to very small SJ
contribution.

In Fig.~16 the dependence of $\bar{p}/p$ ratios at
$\vert y_{c.m.} \vert =0$ is shown as a function of ``centrality"
($\nu$) in $d$Au collisions at $\sqrt{s}=200\,$GeV \cite{PHOB}. The
experimental data are shown here by open squires and the QGSM
predictions with SJ contribution by the solid curve which is very close
to the constant. The reason for such behaviour is that the energy is
high enough, so both the spectrum of $\bar{p}$ and the contribution
of SJ diffusion to the proton spectrum from the target nucleons are
approximately proportional to $\nu$. As a result their ratio is
practically $\nu$-independent. The calculation without SJ contribution
(dashed curve in Fig.~16) is also practically constant, but it again
leads to high values of $\bar{p}/p$ ratios similarly to the case shown
in Fig.~6.  The close points in Fig.~16 present the predictions of the
DPMJET-III model \cite{DPMJET} and they are in agreement with the data,
as well as with QGSM calculations. Dash--dotted curve in Fig.~16 shows
the QGSM predictions for $\bar{\Lambda}/\Lambda$ ratios.

\begin{figure}[h]
\centering
\vskip -0.5cm
\includegraphics[width=.55\hsize]{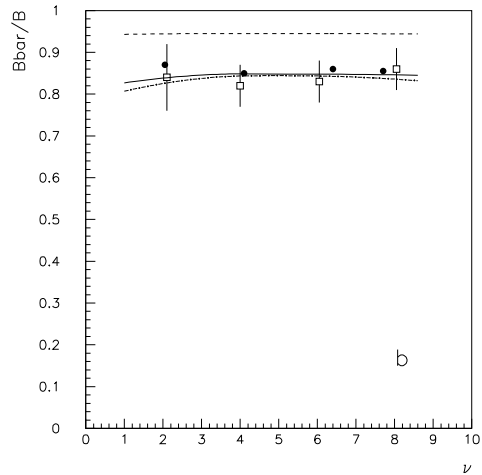}
\vskip -1cm
\caption{\footnotesize
The experimental $\bar{p}/p$ ratio as a function of ``centrality" for
$d$Au collisions at $\sqrt s=200\,$GeV (open squares) together with the
QGSM calculations with SJ (solid curve) and without SJ (dashed curve)
and with the DPMJET-III model (closed points) predictions. The QGSM
predictions for $\bar{\Lambda}/\Lambda$ ratio are shown by dash--dotted
curve}
\end{figure}

The predictions of several another models \cite{HIJING}--\cite{AMPT}
are in some disagreement with the data of \cite{PHOB} (see Fig.~4 in
\cite{PHOB}). The extrapolation of the predictions of these models to
$\nu=1$ give the values of $\bar{p}/p$ in $pp$ interactions larger
than 0.9 that contradicts the data presented both in Figs. 6 and 16.

\section{Conclusion}

We discuss the role of string junction diffusion for the baryon charge
transfer over large rapidity distances for the cases of collisions with
nucleon and nuclear targets. The accounting for the SJ contribution
shown in Fig.~4c with parameters (15) allows one to describe, on a
reasonable level, the main piece of the existing experimental data. The
calculations of the baryon/antibaryon yields and asymmetries without SJ
contribution disagree with the most experimental data, where this 
contribution should be important. The discussed string junction effects 
has $A$-dependences which in general agree with the QGSM predictions 
(see, for example, Figs. 11 and 13).

It is necessary to note that the existing experimental data are not
enough for determination of the SJ parameters with the needed
accuracy. Certain data disagree with the other ones, for example, the  
experimental behaviour of $\bar{\Lambda}$ spectra at $x_F>0$ obtained by
\cite{Mik} and \cite{WA89}, see Table~1, and the experimental 
$\Lambda/\bar{\Lambda}$ asymmetries in \cite{WA89} and \cite{E769} which 
are presented in Figs. 14a and 14b. There exists a disagreement in 
yields of secondary protons with $x_F=0$ produced in $pp$ collisions 
which were measured in \cite{ISR} and \cite{AB}.

We are grateful to G. H. Arakelyan, F.~Bopp, A.~Capella, A.~B.~Kaidalov,
L.~N.~Lipatov, C.~Merino, O.~I.~Piskounova, M.~G.~Ryskin and
A.~A.~Rostovtsev for useful discussions. This paper was supported by
DFG grant GZ: 436~RUS~113/771/1-2 and, in part, by grants
RSGSS-1124.2003.2 and PDD (CP) PST.CLG980287.


\newpage

\begin{thebibliography}{**}

\bibitem{KTM} A.B. Kaidalov, K.A. Ter-Martirosyan, Yad. Fiz. {\bf
39}, 1545 (1984); {\bf 40}, 211 (1984).

\bibitem{KaPi} A.B. Kaidalov, O.I.~Piskunova, Yad. Fiz. {\bf 41},
1278 (1985).

\bibitem{2r} A. Capella, U. Sukhatme, C.I.~Tan, J.~Tran~Thanh~Van,
Phys. Rep. {\bf 236}, 225 (1994).

\bibitem{CaTran} A. Capella, J.~Tran~Thanh~Van, Z. Phys. C~{\bf10},
249 (1981).

\bibitem{KTMS} A.B. Kaidalov, K.A. Ter-Martirosyan and Yu.M.~Shabelski,
Yad. Fiz. {\bf 43}, 1282 (1986).

\bibitem{Sh} Yu.M. Shabelski, Yad. Fiz. {\bf44}, 186 (1986).

\bibitem{Sh1} Yu.M. Shabelski, Nucl. Phys. Proc. Suppl. B~{\bf52}, 116
(1997).

\bibitem{AGK} V.A. Abramovsky, V.N. Gribov and O.V.~Kancheli, Yad. Fiz.
{\bf 18}, 595 (1973).

\bibitem{Kai} A.B. Kaidalov, Sov. J. Nucl. Phys. {\bf 45}, 902 (1987);
Yad. Fiz. {\bf43}, 1282 (1986).

\bibitem{Artru} X. Artru, Nucl. Phys. B {\bf85}, 442 (1975).

\bibitem{IOT} M. Imachi, S. Otsuki and F.~Toyoda, Prog. Theor. Phys.
{\bf 52}, 346 (1974); {\bf 54}, 280 (1976); {\bf 55}, 551 (1976).

\bibitem{RV} G.C. Rossi, G.~Veneziano, Nucl. Phys. B~{\bf123}, 507
(1977).

\bibitem{MRV} L. Montanet, G.C.~Rossi and G.~Veneziano, Phys. Rep.
{\bf63}, 149 (1980).

\bibitem{Khar} D. Kharzeev, Phys. Lett. B~{\bf378}, 238 (1996).

\bibitem{IOT1} M. Imachi, S. Otsuki and F.~Toyoda, Progr. Theor. Phys.
{\bf57}, 517 (1977).

\bibitem{IOT2} M. Imachi, S. Otsuki and F.~Toyoda, Prog. Theor. Phys.
{\bf 52}, 715 (1974).

\bibitem{IOT5} M. Imachi, S. Otsuki and F.~Toyoda, Prog. Theor. Phys.
{\bf55}, 1211 (1976).

\bibitem{KKKY} H. Kanada {\em et al}., Progr. Theor. Phys. {\bf59},
2162 (1978).

\bibitem{Noda} H. Noda, Progr. Theor. Phys. {\bf 68}, 1406 (1982).

\bibitem{ait} E.M. Aitala {\em et al.}, E769 Coll., hep-ex/0009016;
Phys. Lett. B~{\bf469}, 9 (2000).

\bibitem{Bren} A.E. Brenner {\em et al}., Phys. Rev. D~{\bf26}, 1497
(1982).

\bibitem{NA49} H.G. Fischer, NA49 Coll., Nucl. Phys. A~{\bf715}, 118
(2003); hep-ex/0209043.

\bibitem{WA89} M.I. Adamovich {\em et al.}, WA89 Coll., Z. Phys.
C~{\bf76}, 35 (1997); Eur. Phys. J. C~{\bf26}, 357 (2003).

\bibitem{ISR} M. Banner {\em et al.}, Phys. Lett. B~{\bf41}, 547 (1972);\\
B. Alper {\em et al.}, Nucl. Phys. B~{\bf100}, 237 (1975).

\bibitem{BRA} I.G. Bearden {\em et al.}, BRAHMS Coll., Phys. Lett.
B~{\bf607}, 42 (2005);\newline nucl-ex/0409002.

\bibitem{RHIC} B.H. Samset {\em et al}., BRAHMS Coll., Submitted to the
Quark Matter 2004 Int. Conf., Oakland, Jan. 2004.

\bibitem{H1} C. Adloff {\em  et al.}, H1 Coll., Submitted to the 29th
Int.  Conf. on High Energy Physics ICHEP98, Vancouver, July 1998.

\bibitem{KP1} B.Z. Kopeliovich and B.~Povh, Z. Phys. C~{\bf75} (1997) 693.

\bibitem{KP2} B.Z. Kopeliovich and B. Povh, Phys. Lett. B~{\bf446}
(1999) 321.

\bibitem{GKP} G.T. Garvey, B.Z. Kopeliovich and B.~Povh,
Comments Mod. Phys. \newline A~{\bf2}, 47 (2001); hep-ph/0006325.

\bibitem{Bopp} F. Bopp, hep-ph/0002190; hep-ph/0007229.

\bibitem{ACKS} G.H. Arakelyan, A. Capella, A.B.~Kaidalov and
Yu.M.~Shabelski, \newline
Eur. Phys. J. C~{\bf26}, 81 (2002); hep-ph/0103337.

\bibitem{SJ1} F. Bopp and Yu.M. Shabelski, Yad. Fiz. {\bf 68}, 2155
(2005); hep-ph/0406158.

\bibitem{AMS} G.H. Arakelyan, C. Merino and Yu.M.~Shabelski,
Yad. Fiz. {\bf 69}, 911 (2006); hep-ph/0505100.

\bibitem{AMS1} G.H. Arakelyan, C. Merino and Yu.M.~Shabelski,
hep-ph/0604103.

\bibitem{Olga} O.I. Piskounova, Proc. of the HERA-LHC Workshop,
DESY, March 2005.

\bibitem{Sh2} Yu.M. Shabelski, Nucl. Phys. B~{\bf132}, 491 (1978).

\bibitem{CK} A. Capella and A. Krzywicki, Phys. Rev. D~{\bf18}, 3357
(1978).

\bibitem{ASS} V.V. Anisovich, Yu.M. Shabelski and V.M.~Shekhter,
Yad. Fiz. {\bf 28}, 1063 (1978); Nucl. Phys. B~{\bf133}, 477 (1978).

\bibitem{latt} V.G. Bornyanov {\em et al.}, Uspekhi Fiz. Nauk {\bf174},
19 (2004).

\bibitem{Ven} G. Veneziano, Nucl. Phys. B~{\bf117}, 519 (1976).

\bibitem{DPP1} S. Fleck {\em et al.}, Phys. Lett. B {\bf220}, 616
(1989).

\bibitem{DPP2} D. Diakonov, V. Petrov and M.~Polyakov, Z. Phys.
A~{\bf359}, 305 (1997).

\bibitem{DPP3} I.M. Narodetskii, Yad. Fiz. {\bf68}, 780 (2005).

\bibitem{LF} E.M. Levin and L.L. Frankfurt, Pisma v ZhETF {\bf2}, 105
(1965); \\
H.J. Lipkin and F. Scheck, Phys. Rev. Lett. {\bf 16}, 71 (1966).

\bibitem{AKNS} V.V. Anisovich, M.N. Kobrinsky, J.~Nyiri and
Yu.M.~Shabelski, Soviet Physics -- Uspekhi {\bf144}, 553 (1984);
{\em Quark Model and High Energy Collisions}, World Scientific, Singapore,
1985.

\bibitem{MuNa} B.~Muller and J.L.~Nagle, nucl-th/0602029.

\bibitem{AnSh} V.V.~Anisovich and V.M.~Shekhter, Nucl. Phys.
B~{\bf 55}, 455 (1973).

\bibitem{CS} A. Capella and C.-A. Salgado, Phys. Rev. C~{\bf 60},
054906 (1999).

\bibitem{14r} B.Z. Kopeliovich and B.G.~Zakharov, Phys. Lett.
B~{\bf211}, 221 (1988); \\
E. Gotsman and S. Nusinov, Phys. Rev. D~{\bf22}, 624 (1980).

\bibitem{22r} A. Capella and B.Z.~Kopeliovich, Phys. Lett. B~{\bf381}
 325 (1996).

\bibitem{AB} M. Aguilar-Benitez {\em et al.}, LEBC-EHS Coll. Z. Phys.
{\bf50} 405 (1991).


\bibitem{WA97} F. Antinori {\em et al.}, WA97 Coll., Nucl. Phys.
B~{\bf681}, 141 (2001).

\bibitem{SJ2} F. Bopp and Yu.M. Shabelski, Eur. Phys. J. A~{\bf 28},
237 (2006);\newline hep-ph/0603193.

\bibitem{E769} G.A. Alves {\em et al.}, E769 Coll., Phys. Lett
B~{\bf559}, 179 (2003);\newline hep-ex/0303027.

\bibitem{Mik} S. Mikocki {\em et al.}, Phys. Rev. D~{\bf34}, 42 (1986).

\bibitem{PHOB} B.B. Back {\em et al.}, PHOBOS Coll., Phys. Rev.
C~{\bf70}, 011901 (2004); \newline nucl-ex/0309013.

\bibitem{DPMJET} F.W. Bopp, J. Ranft, R.~Engel and S.~Roesler,
hep-ph/0505035.

\bibitem{HIJING} M. Gyulassy and X.N. Wang, Comput. Phys. Commun.
{\bf83}, 307 (1994).

\bibitem{RQMD} H. Sorge, Phys. Rev. C~{\bf52}, 3291 (1995).

\bibitem{AMPT} Z. W. Lin {\em et al}., Phys. Rev. C~{\bf64}, 011902
(2001);\\ B. Zhang {\em et al.}, Phys. Rev. C~{\bf61}, 067901 (2000).


\end{thebibliography}
\end{document}